\documentclass[12pt,letterpaper]{article}
\usepackage{jheppub}
\usepackage{verbatim}

\usepackage{graphicx}
\usepackage{subfigure}
\input{epsf}
\usepackage{epsfig}
\usepackage{epstopdf}
\usepackage{amsthm}

\def\({\left(} \def\){\right)}
\def\[{\left[} \def\]{\right]}
\def\al{\alpha} \def\bt{\beta}
\def\del{{\partial}}

\newcommand{\non}{\nonumber \\}

\newcommand{\be}{\begin{equation}}
\newcommand{\ee}{\end{equation}}
\newcommand{\bea}{\begin{eqnarray}}
\newcommand{\eea}{\end{eqnarray}}
\newcommand{\ba}{\begin{eqnarray}}
\newcommand{\ea}{\end{eqnarray}}

\newcommand{\beq}{\begin{equation}}
\newcommand{\eeq}{\end{equation}}
\newcommand{\beqa}{\begin{eqnarray}}
\newcommand{\eeqa}{\end{eqnarray}}
\newcommand{\beqar}{\begin{eqnarray*}}
\newcommand{\eeqar}{\end{eqnarray*}}

\newcommand{\reef}[1]{(\ref{#1})}

\newcommand{\ie}{{\it i.e.,}\ }

\newcommand{\mt}[1]{\textrm{\tiny #1}}

\def\Xint#1{\mathchoice
   {\XXint\displaystyle\textstyle{#1}}%
   {\XXint\textstyle\scriptstyle{#1}}%
   {\XXint\scriptstyle\scriptscriptstyle{#1}}%
   {\XXint\scriptscriptstyle\scriptscriptstyle{#1}}%
   \!\int}
\def\XXint#1#2#3{{\setbox0=\hbox{$#1{#2#3}{\int}$}
     \vcenter{\hbox{$#2#3$}}\kern-.5\wd0}}

\def\dashint{\Xint-}

\title{Universality of fast quenches from the conformal perturbation theory}

\author[a]{Anatoly Dymarsky}
\emailAdd{a.dymarsky@uky.edu}
\affiliation[a]{Department of Physics and Astronomy, University of Kentucky, \\ Lexington, KY 40506, USA \\
Skolkovo Institute of Science and Technology, Skolkovo Innovation Center, \\ Moscow 143026 Russia}

\author[b]{and Michael Smolkin}
\emailAdd{michael.smolkin@mail.huji.ac.il}
\affiliation[b]{The Racah Institute of Physics, The Hebrew University of Jerusalem, \\ Jerusalem 91904, Israel}

\abstract{
We consider global quantum quenches, a protocol when a continuous field theoretic system in the ground state is driven by a homogeneous time-dependent external interaction. When the typical inverse time scale of the interaction is much larger than all relevant  scales except for the UV-cutoff the system's response  
exhibits universal scaling behavior. We provide both qualitative and quantitative explanations of this universality and argue that physics of the response during and shortly after the quench is governed by the conformal perturbation theory around the UV fixed point. We proceed to calculate the response of one and two-point correlation functions confirming and generalizing universal scalings found previously. Finally, we discuss late time behavior after the quench and argue that all local quantities will equilibrate to their thermal values specified by an excess energy acquired by the system during the quench.  
}

\begin{document}
\maketitle

\section{Introduction}
Quantum quench is a protocol to manipulate quantum system by changing its Hamiltonian while the system at all times follows unitary time evolution. Usually the system before the quench is taken to be in the ground state. Loosely speaking quantum quenches can be divided into two broad types: ``sudden" quantum quenches when the change of Hamiltonian is instantaneous $\hat{H}_0\rightarrow \hat{H}_1$, and the so-called ``smooth" quantum quenches when the Hamiltonian is continuously changed from $\hat H_0$ to $\hat H_1$ over some finite time interval $\delta t$. 

Dynamics of many-body systems following a quantum quench has become an active topic of research recently due to progress with an experimental control of cold atom systems \cite{expquen,kinoshita2006quantum,gring2012relaxation}.  An important question that emerged  in this context is whether an isolated quantum system in a pure non-stationary state can thermalize and be described by some statistical ensemble. This question has received a lot of attention and has been studied intensively both analytically and numerically \cite{polkovnikov2011colloquium,yukalov2011equilibration,quen1,quen2}.

A particular interesting scenario is when a quench occurs near or across a critical point. When the quench is ``slow" in comparison with the gap or other relevant IR scale local observables exhibit  Kibble-Zurek scaling  \cite{KZ,Zurek:1985qw,Dziarm,moore}. In the opposite regime of a sudden, \ie instantaneous  quench when $H_0$ describes a conformal theory Calabrese and Cardy have obtained universal exact results for the two dimensional theories  \cite{Calabrese:2006rx,Calabrese:2007rg};  also see \cite{Delfino:2014qfa,Delfino:2016bln} for recent developments in perturbative formulation of the instantaneous quantum quench problem near criticality in the $1+1$-dimensional case.
In \cite{Buchel:2012gw,Buchel:2013lla,Buchel:2013gba,Das:2014jna,Das:2014hqa,Das:2015jka,Das:2016eao,Das:2017sgp} the authors studied 
an intermediate regime of fast but smooth quenches in a CFT deformed by a relevant operator $\mathcal{O}$. The considered protocol did not assume that the relevant perturbation has to vanish at any point before or during the quench, thus also covering non-conformal theories. Quite remarkably they found that the one-point function of $\mathcal{O}$ during the quench, as well as the excess energy density after the quench, exhibit universal behavior controlled by the conformal dimension $\Delta$ of $\mathcal O$. In the limit when the duration of the quench $\delta t\rightarrow 0$ and  ${d\over 2}\leq \Delta<d$ the response of the system becomes singular, which is at odds with the predictions of sudden quench approximation \cite{Calabrese:2006rx,Calabrese:2007rg,Sotiriadis:2010si,Gritsev:2009wt}. The authors concluded that the fast and sudden protocols represent two different types of processes, with the latter being physically unachievable for field theoretic systems with infinite UV cutoff $a_0^{-1} \rightarrow \infty$.

In this paper we consider continuous relativistic systems undergoing a smooth and spatially-homogeneous quantum quench. Following \cite{Das:2014hqa,Das:2016eao}, we argue that when the quench is fast, i.e.~$\delta t$ is much smaller than all relevant physical scales of the problem except for the UV cutoff, $\delta t\ll a_0$, the behavior of the system during and shortly after the quench is independent of the IR details  and governed by physics at the UV fixed point. Furthermore, when the amplitude of the quench is small compared with an appropriate power of $\delta t$ the system's response can be found in terms of the conformal perturbation theory around the UV CFT. Our findings extend and generalize previous results concerning fast quenches in several important ways. In particular, we put forward a new argument elucidating the mechanism behind the above-mentioned universal behavior. We employ conformal perturbation theory to compute one- and two-point correlation functions of {\it arbitrary} scalar operators to leading order and establish associated universal scalings. Finally, we argue that at late times, beyond the reach of perturbation theory, the system thermalizes in a sense that expectation values of local observables equilibrate to their thermal values.

We start our consideration in section 2 by describing  the wave-function of the system after the quench. We argue that the form of the wave-function justifies conformal perturbation theory and ensures thermalization at late times. This qualitative consideration is followed by an explicit calculation of one- and two-point function during and after the quench  in the conformal perturbation theory. In particular, we calculate time-dependent expectation values for all primary scalars at leading order and for $\mathcal O$ up to second order in perturbation. This is done in section 3. For the two-point functions we discuss different  regimes when the operators in question are time- or spatially-separated. Universal behavior of these correlators is a subject of  section 4. We conclude with a discussion in section 5.

\section{Wave-function and thermalization after the quench}
\label{WF}
We start our consideration by describing the evolution of system's wave-function following a quantum quench. We assume that the system in question is described by a conformal field theory in the UV, while the details of the  IR behavior are not important. The UV cutoff and IR scale where the UV CFT description breaks down will be denoted as $a_0^{-1}$ and $m$ respectively. For the sake of simplicity in this section we assume that before and after the quench the system is described by the same Hamiltonian $H_0$, 
\bea
\label{Hamlt}
H(t)=H_0+\lambda(t)\int d^{d-1}x\, \mathcal O\ ,
\eea
where the deformation $\mathcal O$ is relevant in the UV and has scaling dimension $\Delta<d$. The  time-dependent coupling 
\beq
 \lambda(t)=\delta\lambda \, f(t/\delta t)
 \label{profile}
\eeq
is a ``pulse" of amplitude $\delta \lambda\propto\ell^{\Delta -d}$, and
$f(t)$ is a dimensionless smooth function which vanishes sufficiently fast outside some interval of order one centered around $t=0$.  Since $\lambda(t)$ approaches zero at infinity it has a well-defined Fourier transform which we denote as $\tilde{\lambda}(\omega)$.  A more general case when $\lambda(t)$ asymptotes to a non-zero constant is delegated to Appendix \ref{trans}. 

Before the quench the system is in the ground state $|0\rangle$ of $H_0$.  Time dependent wave-function can be decomposed in terms of energy eigenstates of $H_0$, $|\Psi(t)\rangle=\sum a_n(t) |n\rangle$. Expanding in $\delta \lambda$, at leading order in perturbation theory the probability of vacuum decay is given by (see Appendix \ref{trans})
\bea
\label{sumP}
P=\lim_{t\rightarrow \infty}\sum_{n\neq 0} |a_n|^2=(2\pi)^{d-1} V \sum_{n\neq 0}
|\tilde{\lambda}(E_n)|^2 \delta(\vec{p}_n) \left|\langle 0|\mathcal O(0)|n\rangle\right|^2\ .
\eea
Here $\vec{p}_n$ and $E_n$ is the momentum and energy of the state $|n\rangle$. Clearly, because of the translational invariance of \eqref{Hamlt} only states with zero momentum contribute to $|\Psi(t)\rangle$. The sum in \eqref{sumP} can be conveniently expressed in terms of the spectral density function
\bea
\label{rhodef}
{\rho(p^2)\Theta(p^0)\over (2\pi)^{d-1}}=\sum_n \delta^d(p-p_n) \left|\langle 0|\mathcal{O}(0)|n\rangle\right|^2 ~,
\eea
that controls imaginary part of $\langle 0|{\mathcal O}(p) {\mathcal O}(-p)|0\rangle$ via K$\ddot{\rm a}$ll$\acute{\rm e}$n-Lehmann representation. Combining \eqref{sumP} and \eqref{rhodef} gives probability density 
\bea
\label{probability}
\mathcal{P}={P\over V}=\int_{0^+}^\infty d\omega\, \rho(\omega^2) |{\lambda}(\omega)|^2\ .
\eea 

Local operator $\mathcal O$ acting on vacuum can only create localized states, which can be colloquially thought of as a cluster of particles. Transitionally-invariant states $|n\rangle$ should be understood as these localized states smeared over the entire space. 
Probability density ${\mathcal P}$ then has an interpreted of $dP/dV$, the probability of creating a localized excited state within a unit volume of space during the quench. Creation of particles at different locations are independent random processes, with the full number of excited states created within any given volume given by the Poisson distribution. More precisely \reef{probability} describes creation of states with different energies, which at leading order in perturbation theory are independent Poisson processes.  Accordingly, average energy density produced during the quench will be given by 
\bea
\label{density}
{\epsilon}={\bar E\over V}=\lim_{t\rightarrow \infty}{\langle \Psi|H_0|\Psi\rangle\over V}=\int_{0^+}^\infty d\omega\,\omega\, \rho(\omega^2) |{\lambda}(\omega)|^2\ ,
\eea
and energy  fluctuations by 
\bea
\label{fluctuations}
{\Delta E^2\over V}= \lim_{t\rightarrow \infty}{\langle \Psi|H^2_0|\Psi\rangle-\langle \Psi|H_0|\Psi\rangle^2\over V}=\int_{0^+}^\infty d\omega\,\omega^2\, \rho(\omega^2) |{\lambda}(\omega)|^2\ .
\eea

Spectral density function $\rho(\omega^2)$ is not known in general. But for large $\omega\gg m$ it can be deduced from the short-distance behavior of $\langle 0|{\mathcal O}(x){\mathcal O}(0)|0\rangle$, which is fixed by the conformal symmetry in the UV  (see Appendix \ref{cftspec}),
\bea
\rho(\omega^2)\propto\omega^{2\Delta-d}\ .
\label{speconf}
\eea 
In the case of CFT equation \reef{density} can be also derived using the standard Ward identity for the stress-tensor,  see Appendix \ref{Ward} for details.

For $\Delta>d/2$, the integrand in \eqref{probability} grows with $\omega$. When $\lambda(t)$ is a smooth function, e.g.~$\lambda(t)\propto e^{-(t/\delta t)^2}$, its shape provides a dynamical upper cutoff at $\omega\sim \delta t^{-1}$, where the integral is saturated. This means typical localized states produced during the quench will have energy $E\sim \delta t^{-1}$. The average energy density and energy fluctuations follow from (\ref{density}, \ref{fluctuations})  
\bea
\label{DeltaE}
{\epsilon} \sim \ell^{2\Delta -2d}\, \delta t^{d-2\Delta}\ ,\qquad \Delta E^2 \sim V \ell^{2\Delta -2d}\, \delta t^{d-2\Delta-1}\ .
\eea

In the discussion above we implicitly assumed that the resulting density of particles after the quench is not too large such that the probability of particles to interact during the quench is small. To justify this assumption we consider volume $\Delta V=\Delta L^{d-1}$ such that the probability of finding a localized state inside $\Delta V$ after the quench  is of order one, 
\bea
{\mathcal P}\Delta V\sim 1\ ,\qquad  \Delta L\sim \delta t \left({\ell\over \delta t}\right)^{2(d-\Delta)/(d-1)}\ .
\eea
Here $\Delta L$ is a typical distance between two neighboring localized states and due to causality these states would not be able to interact during the quench so far 
\bea
\label{scales}
\ell\gg \delta t\ ,
\eea
and consequently $\Delta L\gg \delta t$. The same condition  ensures that Compton wavelength of the created particles $\delta t$ is much smaller than the distance between them $\Delta L$, which justifies treatment of these states as fully localized. 
 
When $m^{-1}\gg \delta t \gg a_0$ and $\ell \gg \delta t$, qualitative time evolution of the system can be summarized as follows. During the quench ground state decays into a diluted ``gas" of highly excited localized states each of an approximate energy $\delta t^{-1}$.  Typical distance between these states $\Delta L$ is large enough such that 
up to time $t\sim \Delta L$ these states do not interact with each other. This specifies the limits of applicability of the conformal perturbation theory. Up to the times 
\bea
\label{maxtime}
t\lesssim \delta t \left({\ell\over \delta t}\right)^{2(d-\Delta)/(d-1)}
\eea
dynamics of the system can be described in terms of the UV CFT, but once the localized excitations start colliding, perturbative approximation breaks down and the system equilibrates. Although beyond this point we can not describe system's dynamics in detail, energy distribution (\ref{density}), (\ref{fluctuations}) will remain the same so far Hamiltonian is time-independent. Accordingly, \eqref{fluctuations} shows that at late times energy fluctuations $\Delta E$ will scale with the volume as $V^{1/2}$. Combined with the standard argument of eigenstate thermalization this means local physics will be thermal, with the effective temperature determined by energy density \eqref{density},
\bea
\label{T}
T\sim \delta t^{-1} (\delta t/\ell)^{2(d-\Delta)/d}\ .
\eea
As a consistency condition \eqref{T} requires the effective temperature to be high in comparison with the IR scale, $T\gg m$, which introduces a constraint on $\ell$ in terms of $\delta t$ and $m$.

In other words, our qualitative description of $|\Psi(t)\rangle$ allows us to make a highly non-trivial prediction that following the quantum quench system will thermalize in the sense that local observables will equilibrate to their thermal values. Let us emphasize here that the system always remains in a pure state and never approaches true thermal ensemble. This can be seen already from the expression for $\Delta E$ \eqref{DeltaE}, which  is different from the energy fluctuations for the conformal field theory in a Gibbs state,  $\Delta E^2\sim V T^{d+1}$. Accordingly, the prediction of thermalization can be only extended to local quantities, confined to the subregion much smaller than the full volume $V$.

In conclusion we discuss the limit when the fast and smooth quench approaches the sudden one, $\delta t\sim a_0$. When the quench is truly instantaneous with $\lambda(t)$ given by the step-function\footnote{To show that the divergence of \eqref{probability} is a result of an abrupt change of $\lambda$ and is not related to its asymptotic behavior at large $t$, one can consider a rectangular-shaped pulse with $\tilde{\lambda}\sim \sin(\omega \delta t)/\omega$, which would suffer from the same divergence.} $\tilde\lambda(\omega)\sim \omega^{-1}$, the probability integral \eqref{probability} diverges for $\Delta>(d+1)/2$. This means the actual upper limit in \eqref{probability} will be given by the physical UV cutoff $a_0$, i.e.~during the quench vacuum decays into some highly excited states  of energy $a_0^{-1}$. These excitations can not be described in terms of the UV CFT and the resulting state has no universality, i.e.~it is sensitive to the details of UV physics. This explains the very different physical behavior observed for sudden and fast quenches  in  \cite{Buchel:2012gw,Buchel:2013lla,Buchel:2013gba,Das:2014jna,Das:2014hqa,Das:2015jka}.

\subsection{Path integral formalism}

The picture described in the previous section, which invoked particles and localized states, may look too qualitative to be precise. In what follows we reproduce main results of the previous section using a non-petrubative path integral formalism. 

Let us consider a field theoretic system governed by the time independent Hamiltonian, $H_\text{in}$. We assume that the system resides in the ground state, $|0, \text{in}\rangle$ when $H_\text{in}$ starts to experience a continuous homogeneous deformation (smooth homogeneous quench) leading to a time independent Hamiltonian $H_\text{out}$ in the future
\be
 H_\text{in} \longrightarrow  H(t)  ~\underset{t\to\infty}{\longrightarrow} ~   H_\text{out}~.
\ee
The coupling $\lambda(t)$ in \reef{Hamlt} is an interpolating parameter between the Hamiltonians. For example, if $H_\text{in}$ exhibits conformal symmetry while $\lambda(t)$ vanishes asymptotically,  this a setup representing a smooth CFT-to-CFT quantum quench with $H_\text{in}=H_\text{out}=H_0$. Let us emphasize though that our discussion is more general and asymptotic Hamiltonians do not have to be equal. 

The quantum quench results in a non-trivial transition matrix, $S$, between the eigenstates $|m, \text{in}\rangle$ and $|n, \text{out}\rangle$ of the momentum operators $P^\mu_\text{in}=(H_\text{in}, \vec P)$ and $P^\mu_\text{out}=(H_\text{out}, \vec P)$ respectively. In our case spatial momentum is conserved and therefore $S$ may be written  as follows
\be
 S_{nm} \equiv \langle n,\text{out}| m, \text{in}\rangle = \delta_{nm} + i \, (2\pi)^{d-1} \, \delta(\vec p_n - \vec p_m) \, T_{nm} ~,
 \label{Smatrix}
\ee
where $\delta_{nm}$ is an ``identity matrix" mapping energy eigenstates of $H_\text{in}$ and  $H_\text{out}$. It describes an adiabatic i.e.~infinity slow transition of $H_\text{in}$ into $H_\text{out}$ when the non-trivial part $T_{nm}$ vanishes.

Unitarity of $S$ requires 
\be
 \text{Im} ~ T_{mm} = {(2\pi)^{d-1} \over 2} \int dn    ~ \delta(\vec p_n - \vec p_m) ~ |T_{nm}|^2 ~,
\label{OT}
\ee
where the integral runs over continuum of ``out" states. This relation is basically a version of the optical theorem.
In  \eqref{OT} we introduced the measure over the out-states 
\be
 dn =  {d\omega \, d^{d-1} \vec p \over (2\pi)^{d-1}} \,\Omega(\omega,  p) ~.
\ee
with help of the properly normalized density of states $\Omega(\omega,  p)$.

In this paper we focus on the initial state $| m, \text{in}\rangle=|0, \text{in}\rangle$, therefore  in \eqref{OT} we have $\vec p_m=0$ and\footnote{Besides the total energy $\omega$ and momentum $\vec p$, the final state $|n,\text{out}\rangle$ may have many other quantum numbers. Matrix element $T_{\omega0}$ is defined such that
$\Omega(\omega,0)|T_{\omega0}|^2$  includes the sum over these quantum numbers.}
\be
 \text{Im} ~ T_{00} = {(2\pi)^{d-1} \over 2} \int_0^\infty d\omega  \int {d^{d-1} \vec p  \over (2\pi)^{d-1}} ~ \Omega(\omega,  p )~ \delta( \vec p ) ~ |T_{\omega 0}|^2 
 ={1 \over 2} \int_0^\infty d\omega  ~ \Omega(\omega ,  0) ~ |T_{\omega0}|^2 ~.
 \label{unitarity}
\ee 

The  probability of the system to be in a vicinity of an excited state $|\omega, \text{out}\rangle\neq |0, \text{out}\rangle$ is given by  
\be
 dP_{\,|0, \text{in}\rangle \to |\omega, \text{out}\rangle} =   {d^{d-1} \vec p  d\omega  \over (2\pi)^{d-1}} ~ \Omega(\omega ,  p )~ |S_{\omega 0}|^2 =  
 V \, d^{d-1} \vec p  \, d\omega  \, \Omega(\omega ,  p ) \, \delta(\vec p )~ |T_{\omega 0}|^2 ~,
\ee
where $V$ is the full volume of space
\be
 V\equiv(2\pi)^{d-1} \delta (\vec p )\big|_{\vec p =0} = \int d^{d-1} \vec x ~ e^{i\vec p  \cdot \vec x} \Big|_{\vec p =0} ~.
 \label{vol}
\ee
Now we see that the right hand side of \reef{unitarity} (up to an overall volume factor) has simple interpretation in terms of the total probability for the vacuum to decay into states $|0, \text{in}\rangle \to |\omega, \text{out}\rangle$ during the quench\footnote{Integral on the right hand side of \reef{unitarity} includes ground state. However, it should be excluded from the probability of vacuum decay, therefore in \reef{transP} we split the integral into two parts:  constant contribution of the vacuum state and integral over the excited states only (dashed integral stands to remind that vacuum is excluded). } 
\be
V\, \text{Im} ~ T_{00}  ={1 \over 2} \dashint_{0}^\infty d\omega  ~ \Omega(\omega ,  0) ~ |T_{\omega 0}|^2 +{V^2\over 2}|T_{00}|^2 = {1 \over 2} \int d{P}_{\,|0, \text{in}\rangle \to |\omega, \text{out}\rangle} + {V^2\over 2}|T_{00}|^2~.
  \label{transP}
\ee

The transition amplitude   $T_{00}$ is given by the sum of all {\it connected} and {\it disconnected} vacuum diagrams with at least one insertion of $\mathcal{O}$. With help of \reef{Smatrix} it can be easily  related to the effective action $\Gamma_\mt{eff}[\lambda]$, which is a sum of only {\it connected} diagrams
\be
\exp(i\Gamma_\text{eff})\equiv \langle 0, \text{out}|0, \text{in} \rangle= S_{00}=1 + i \, V \, T_{00} ~.
\label{Geff}
\ee
In perturbation theory we expand $T_{00}=\sum_{j=1}^\infty T_{00}^{(j)}$, where $T_{00}^{(j)}$ is proportional to $\delta \lambda^j$.
Up to second order in $\delta \lambda$ we find 
\bea
 V T_{00}^{(1)}& =& - \int_{-\infty}^\infty dt\, d^{d-1} \vec{x}\,  \lambda(t) \langle0, \text{in}| \mathcal{O}(t,\vec x) |0, \text{in}\rangle ~,
 \label{T1}\\
 V T_{00}^{(2)}&=& {i\over 2} \int_{-\infty}^\infty dt_1 \lambda(t_1)\int_{-\infty}^\infty dt_2 \lambda(t_2)   ~
 \langle0, \text{in}| \mathcal{T} \Big( \mathcal{O}(t_1)\mathcal{O}(t_2) \Big)|0, \text{in}\rangle~,
  \nonumber
\eea
where in the second line the correlator is time-ordered, and we introduced shorthand notation $\mathcal{O}(t)\equiv \int d^{d-1}\vec x \, \mathcal{O}(t,\vec x)$.
In general $\langle0, \text{in}|\mathcal{O}|0, \text{in}\rangle$ is non-vanishing and real. Hence, combining \reef{transP} and  \reef{Geff} we have to second order in $\delta\lambda$
\be
 2\,\text{Im}\,\Gamma_\mt{eff}^{(2)} =2 V\text{Im}\, T^{(2)}_{00} -V^2 \big(T^{(1)}_{00}\big)^2=\int d{P}_{\,|0, \text{in}\rangle \to |\omega, \text{out}\rangle} ~.
 \label{Geff2}
\ee
On the other hand, from \reef{T1}
\bea
2\,\text{Im}\,\Gamma_\mt{eff}^{(2)}
=\text{Re}  \int_{-\infty}^\infty dt_1 \lambda(t_1)\int_{-\infty}^\infty dt_2 \lambda(t_2)   ~
 \langle0, \text{in}| \mathcal{T} \Big( \mathcal{O}(t_1)\mathcal{O}(t_2) \Big)|0, \text{in}\rangle_{\rm con.} 
\label{qwe}
\eea
We see that $\big(T^{(1)}_{00}\big)^2$ subtracts the disconnected part from the correlator of two $\mathcal{O}$'s leaving the connected piece. Using K$\ddot{\rm a}$ll$\acute{\rm e}$n-Lehmann representation  
\be 
\label{KLR}
\langle0, \text{in}| \mathcal{T} \Big( \mathcal{O}(t_1,\vec x_1)\mathcal{O}(t_2,\vec x_2) \Big)|0, \text{in}\rangle_{\rm con.} = -i
 \dashint_{0}^\infty d\mu^2 \rho(\mu^2)\int {d^d p\over (2\pi)^d} {e^{i\,p\cdot(x_1-x_2)} \over p^2+\mu^2-i\epsilon}~.
\ee
Integrating over $t_1, t_2, \vec x_1$ and $\vec x_2$ gives
\be
2\,\text{Im}\,\Gamma_\mt{eff}^{(2)} ={V} \, \text{Im} \dashint_{0}^\infty d\mu^2 \rho(\mu^2) \int_{-\infty}^{\infty} {d\omega\over 2\pi} ~ {|\lambda(\omega)|^2\over -\omega^2 + \mu^2-i\epsilon} ~.
\ee
Using \eqref{Geff2} and the standard identity 
\be
\text{Im} \, {(-\omega^2 + \mu^2-i\epsilon)^{-1}} =  \pi \delta( \mu^2 - \omega^2) ~,
\ee
we recover \reef{probability}, see also Appendix \ref{trans}.

We notice that both \reef{density} and \reef{fluctuations} are given by the connected piece of the two-point function \eqref{KLR}. This is a general rule, at all orders in perturbation theory the momenta of energy distributions $\langle H \rangle_\text{out}$, 
$\langle\langle H^2\rangle\rangle_\text{out}=\langle H^2\rangle_\text{out}-\langle H \rangle_\text{out}^2$, etc., will be given by the connected diagrams contributing to the effective action \reef{Geff}. In particular, to calculate the $k$-th moment of energy distribution $\langle\langle H^k\rangle\rangle_\text{out}$ to second order in $\lambda$, it is enough to multiply the integrand in \reef{probability} by $\omega^k$. Thus, for instance, if $\Delta>d/2$ and $\delta t$ satisfies \reef{scales}, one can replace $\rho(\omega^2)$ with its conformal counterpart \reef{speconf} to find
 \bea
  \langle\langle H^k\rangle\rangle_\text{out} &=& {(4 \pi)^{d\over 2} N \over 2^{2\Delta} \Gamma(\Delta)\Gamma\( 1+\Delta-{d \over 2} \)}
 {\delta\lambda^2\over\delta t^{2\Delta-d+k-1}} \int_{0}^\infty d\tilde\omega ~ {|{ f}(\tilde\omega)|^2 \over \tilde\omega^{d- 2\Delta-k}}\Big(1+ h\, \delta\lambda\delta t^{d-\Delta} +  \ldots\Big)~,
 \nonumber
\eea
where $\tilde\omega=\omega\,\delta t$ is a dimensionless energy, $h$ is some constant, whereas ellipsis stand for higher order terms in $\delta\lambda\,\delta t^{d-\Delta}\ll 1$ and IR scale $m\,\delta t \ll 1$. It follows that higher order corrections  are suppressed  in the limit $\delta t \to 0$ justifying  the conformal perturbation theory around the UV fixed point. Note also that for any $\Delta$ satisfying the unitarity bound the $k$-th moment of energy distribution for $k>3$ exhibits divergence in the limit $\delta t\to 0$.

\section{Quenched one-point functions}
As argued in the previous section, a field theoretic system subject to a fast quench can be described in terms of an appropriate UV CFT. In this section we use conformal perturbation theory to calculate one-point function of arbitrary scalar operators following the quench. Previous works employing conformal perturbation theory in a similar context include \cite{Berenstein:2014cia,Berenstein:2016avf}.  We assume that the microscopic theory \reef{Hamlt} is a deformation of a conformal theory $H_0=H_\mt{CFT}$, and $\lambda(t)$ vanishes at infinity, thus describing a CFT-to-CFT quench. 

Using Schwinger-Keldysh we expand time-dependent wave-function of the system up to second order in $\delta\lambda$, 
\bea
 |\Psi(t)\rangle = e^{-iH_\mt{CFT}(t-t')} \Bigg( &&|\Psi(t')\rangle +
 (- i) \int_{t'}^t d t_1 \, \lambda( t_1)\mathcal{O}(t_1)  |\Psi(t')\rangle 
  \label{state}
 \\
 && + (- i)^2 \int_{t'}^t d t_1 \, \lambda( t_1) \int_{t'}^{t_1} d t_2 \, \lambda( t_2) \mathcal{O}(t_1)\mathcal{O}(t_2)  |\Psi(t')\rangle +\ldots \Bigg)~,
  \nonumber
  \eea
where $\mathcal{O}(t)=\int d^{d-1}\vec x \, \mathcal{O}(t, \vec x)$ with $\mathcal{O}(t,\vec{x})=e^{iH_\mt{CFT}(t- t')} \, \mathcal{O}(\vec x) \, e^{-iH_\mt{CFT}( t-t')}$ being the standard Heisenberg operator in the unperturbed CFT. The state within the parenthesis is what we usually call the ``interaction picture" state, while $ |\Psi(t)\rangle$ is the so-called ``in" state. In what follows we are going to calculate  $\langle  \mathcal{O} \rangle$, where the expectation value is taken in the ``in" state.  Of course, the hierarchy of scales discussed in the previous section, $a_0\ll \delta t\ll \ell$, is implicitly assumed throughout this section as well to ensure validity of the conformal perturbation theory. 

We derive the universal scaling of $\langle  \mathcal{O} \rangle$ in the limit of fast but smooth quenches $\delta t \to 0$ confirming and generalizing previous holographic and free field theory calculations  \cite{Buchel:2013lla,Buchel:2013gba,Das:2014jna,Das:2014hqa,Das:2015jka}. The logarithmic enhancement  found in holography \cite{Buchel:2012gw,Buchel:2013lla,Buchel:2013gba} and free field theories \cite{Das:2014jna,Das:2014hqa,Das:2015jka} for special values of $d$ and $\Delta$ are shown to hold for any CFT and the overall constant factor is explicitly evaluated.\footnote{See also Appendix A in the recent publication \cite{Das:2017sgp} where the authors carry out some of the calculations presented in this section.}

\subsection{First order}
For a CFT starting in the vacuum state before quench at linear order in $\delta \lambda$
\beq
 \langle \mathcal{O}(t,\vec x) \rangle = \langle 0| \mathcal{O}(t,\vec x)|0\rangle ~ - i \int_{-\infty}^t dt' \int d^{d-1}\vec y \, 
 \lambda(t')  \langle 0| [\mathcal{O}(t, \vec x), \mathcal{O}(t',\vec y) ] |0\rangle + \mathcal{O}(\delta\lambda^2) ~,
 \label{1ptO}
\eeq
where we used \reef{state} with $t'=-\infty$, assuming the initial state is the vacuum state of the unperturbed CFT, $|\Psi(-\infty)\rangle=|0\rangle$. Thus, the correlators on the right hand side of \eqref{1ptO} are evaluated in the unperturbed CFT. 

Using translational invariance the integral over space can be evaluated explicitly. We relegate the details of the calculation to Appendix \ref{intcom} (see  \reef{icomm} there), while here present the final answer   
\beq
 \int d^{d-1}\vec x \, \langle 0| [\mathcal{O}(t, \vec x), \mathcal{O}(0,0) ] |0\rangle = -  iN
 {2 \pi^{d+1\over 2} \over \Gamma\big( {d-2\Delta+1\over 2} \big)\Gamma(\Delta)} ~ \text{sign}(t) \,|t|^{d-2\Delta-1}~.
\label{33}
\eeq
Here $N$ is the normalization constant of the Euclidean two-point function of $\mathcal{O}$. With help of \eqref{33} and $\langle 0| \mathcal{O}(x)|0\rangle=0$ we obtain for the linear correction
\beq
 \delta^{(1)}\langle \mathcal{O}(t,\vec x)\rangle =  {-2 \pi^{d+1\over 2} N \over  \Gamma(\Delta) \Gamma\big( {d-2\Delta+1\over 2} \big)} \delta\lambda\int_{-\infty}^t dt'
 \,{f(t'/\delta t) \over (t-t')^{2\Delta-d+1}  }  ~.
  \label{1pt}
\eeq
This expression exhibits a UV divergence for $d/2\leq\Delta< d$.  It emerges at the upper bound $t'=t$ where the two operators $\mathcal{O}(x)$ in the commutator collide. We regulate the integral by introducing a sharp UV cut off $a_0$, \ie the integral runs over $t'$ from $-\infty$ to $t-a_0$. Then to calculate the divergent terms one has to Taylor expand $h(t'/\delta t)$ in the vicinity of $t'=t$ and carry out the divergent integrals. These  divergences are ought to be canceled by adding appropriate counterterms. We note that  if $t$ lies outside the support of $\lambda(t)$, $\delta^{(1)}\langle \mathcal{O}\rangle$ is finite and  counterterms are not necessary. From now on we only consider the scheme-independent logarithmic divergence. This can be achieved, for instance, by employing dimensional regularization or analytically continue in $\Delta$.

When no logarithmic divergence is present, it follows from \reef{1pt} that the one-point function exhibit the following scaling  \cite{Buchel:2013lla,Buchel:2013gba,Das:2014jna,Das:2014hqa,Das:2017sgp},
\be
  \delta^{(1)}\langle \mathcal{O}(t,\vec x)\rangle = \left\{\begin{matrix} 
      &    a_1(t) \delta\lambda\delta t^{d-2\Delta}  ~, \qquad\qquad~          t\in\text{supp}\big(\lambda(t)\big),
      \\
      \\
      &    {\delta\lambda \delta t\over t^{2\Delta-d+1}}\( b_1 + b_2 {\delta t\over t} + \ldots\)
      ~, \quad \delta t\ll t\ll \ell ~. 
\end{matrix} \right.
\label{1stO}
\ee
Here $a_1(t)$, $b_1$ and $b_2$ are of order one and fixed by \reef{1pt}. In the limit $\delta t\rightarrow 0$ \eqref{1stO} is singular for $\Delta>d/2$, and as argued in \cite{Buchel:2013lla,Buchel:2013gba,Das:2014jna,Das:2014hqa}, it suggests that instantaneous process, $\delta t=0$, can not actually be realized. This singularity is clearly at odds with the``sudden quench" approximation of \cite{Calabrese:2006rx,Calabrese:2007rg,Sotiriadis:2010si} when the wave-function does not change which must result in $\langle \mathcal{O}\rangle=0$ across the quench. The crucial parameter which separates these two scenarios is the ratio $\delta t/a_0$. When the latter is small, the dynamics is well described by the UV CFT, but when $\delta t$ is smaller than $a_0$, \eqref{1stO} is no longer applicable. In case of lattice models  this transition can be studied explicitly \cite{Das:2017sgp}.

When \reef{1pt} exhibits a logarithmic divergence, the expression for $\delta^{(1)}\langle \mathcal{O}\rangle$ is local, and for $t$ within the support of $\lambda(t)$ the scaling receives an additional logarithmic enhancement.
These logarithmic terms show up for $\Delta=(d-1+n)/2$ with $n=1,2,..,d+1$, where the upper bound on $n$ is fixed by the requirement $\Delta \leq d$. In this case
\be
 \delta^{(1)}\langle \mathcal{O}(t,\vec x)\rangle\Big|_{\Delta={d-1+n\over2}} =  {2 \pi^{d+1\over 2} N \over  \Gamma\big({d-1+n\over 2}\big) \Gamma\big( {2-n\over 2} \big) \Gamma(n)} {d^{n-1}\over dt^{n-1}}\lambda(t) 
  \log(a_0/\delta t) + \ldots \,, ~ n=1,2,..,d+1 ~,
    \label{log1pt}
\ee
where ellipsis stand for non-universal finite terms. This expression can be further simplified. When $n$ is even and $n\geq 2$, poles of  gamma function in the denominator cancel the log term. These poles are equivalent to $\log(1/a_0)$ in the dimensional regularization scheme. Thus for even $n$ there is precise cancellation between these zeros and logarithmic divergence of the integral over time, and we get the following exact and local expressions\footnote{Note that for even $n$ eq.~\reef{fin1pt} gives an exact answer, while for odd $n$ we suppressed a finite non-local term since it can be changed by rescaling $a_0$, and therefore is non-universal.}
 \bea
  &&\delta^{(1)}\langle \mathcal{O}(t,\vec x)\rangle\Big|_{\Delta={d-1+n\over2}}=
   - {  2\pi^{d-1\over 2} N \Gamma({n\over 2}) \over  \Gamma\big({d-1+n\over 2}\big)  \Gamma(n)} {d^{n-1}\over dt^{n-1}}\lambda(t)
   \nonumber
   \\
 &&\quad\quad\times \left\{\begin{matrix} 
      &  (-1)^{n\over 2}  {\pi \over  2}  \quad\quad\quad\quad\quad~\, , \, \text{for {\bf even}} ~ n\in 2,4,..,d+1 ~,
      \\
      \\
      &   (-1)^{n+1\over 2} \log\big(a_0/ \delta t\big) \quad , \, \text{for {\bf odd} } ~ n\in 1,3,..,d+1 ~.
 \end{matrix} \right.
   \label{fin1pt}
\eea
The cut off dependence inside $\log$ is eliminated by an appropriate counterterm.
Now we clearly see the logarithmic enhancement $\log\delta t$  relative to the naive scaling $\delta t^{d-2\Delta}$ for even integer $n=2\Delta -d+1$.

It is instructive to compare  our results with \cite{Das:2014jna,Das:2014hqa}, which were considering free field theories with the time-modulated mass.  In the case of free scalar, $\Delta=d-2$ (or equivalently, $n=d-3$), $\lambda(t)=m^2(t)$ and $\mathcal{O}=\phi^2/2$. Hence,
\be
 \langle 0| \mathcal{O}(x) \mathcal{O}(0)|0\rangle = {1\over 4}  \langle 0| \phi^2(x) \phi^2(0)|0\rangle 
 ={1\over 2}\langle 0| \phi(x) \phi(0)|0\rangle^2 =  {N_\phi\over x^{2(d-2)}}~,
\ee
where $N_\phi= {\Gamma\big( {d-2\over 2} \big)^2\over 32\pi^d}$. Substituting into \reef{fin1pt}, we find
\be
\delta^{(1)} \langle \phi^2(t,\vec x)\rangle = 
\left\{\begin{matrix} 
      &     {4(-1)^{d/2}\over (16\pi)^{d-1\over 2} \, \Gamma\big( {d-1\over 2} \big)} \partial_t^{d-4} m^2(t) \log(a_0/\delta t)+ \ldots \quad \text{for even } d\geq 4~,
      \\
      \\
      &   {2\pi(-1)^{d-1\over 2}\over (16\pi)^{d-1\over 2} \, \Gamma\big( {d-1\over 2} \big)} \partial_t^{d-4} m^2(t)  
      \quad\quad\quad\quad\quad\quad\quad\quad \text{for odd } d\geq 5~.
\end{matrix} \right.
\ee
These results agree with \cite{Das:2014jna,Das:2014hqa}.

For free fermions $\Delta=n=d-1$, $\lambda(t)=m(t)$ and $\mathcal{O}=\bar\psi\psi$. Accordingly
\be
 \langle 0| \mathcal{O}(x) \mathcal{O}(0)|0\rangle =   \langle 0| \bar\psi\psi(x) \bar\psi\psi(0)|0\rangle 
 = {N_\psi\over x^{2(d-2)}}~,
\ee
where $N_\psi= 2^{[{d\over 2}]}\,{\Gamma\big( {d\over 2} \big)^2\over 4\pi^d}$. Substituting into \reef{fin1pt}, we find
\be
\delta^{(1)} \langle \bar\psi\psi(t,\vec x)\rangle = 
\left\{\begin{matrix} 
      &   2^{{d\over 2}+1}   { (-1)^{d+2\over 2}\over (16\pi)^{d-1\over 2} \, \Gamma\big( {d-1\over 2} \big)} \partial_t^{d-2} m(t) \log(a_0/\delta t)+ \ldots \quad \text{for even } d\geq 4~,
      \\
      \\
      &   2^{[{d\over 2}]}  {\pi(-1)^{d+1\over 2}\over (16\pi)^{d-1\over 2} \, \Gamma\big( {d-1\over 2} \big)} \partial_t^{d-2} m(t)  
      \quad\quad\quad\quad\quad\quad\quad\quad \text{for odd } d\geq 3~.
\end{matrix} \right.
\ee
Again we find agreement with \cite{Das:2014hqa}.\footnote{Up to an overall sign our results match eqs. (3.14), (3.15) in \cite{Das:2014hqa} provided that $(2\pi)^{d-1\over 2} \to (2\pi)^{d-1}$ in eq. (3.10) of that paper. We thank authors of \cite{Das:2014hqa} for correspondence on this matter.}

\subsection{Second order}

Let us now study $\langle\mathcal{O}_i\rangle$, the one-point function of a scalar primary of arbitrary conformal dimension $\Delta_i \neq \Delta,\ \Delta_i<d$. In this case Euclidean correlator $\langle\mathcal{O}\mathcal{O}_i\rangle$ vanishes, thus there is no linear correction to $\langle\mathcal{O}_i\rangle$, and the leading response is  quadratic in $\delta \lambda$.  Using \reef{state} we get after setting $t'=-\infty$ and performing simple algebra
\be
  \delta^{(2)} \langle\mathcal{O}_i(t,\vec x)\rangle=2 \, \text{Re} \int_{-\infty}^t d t_1 \lambda( t_1) \int _{-\infty}^{t_1} dt_2 \, \lambda( t_2) 
~ \langle 0| \Big[\mathcal{O}(t_1),\mathcal{O}_i(t,\vec x)\Big]\mathcal{O}(t_2)|0\rangle ~.
     \label{1ptO2}
\ee

This ordered correlator can be evaluated by assigning an appropriate $i\epsilon$ prescription to the Euclidean three point function. As a general rule, to enforce right ordering of the operators, an operator that is to the ``left" of another should have algebraically smaller negative imaginary part in the time direction. 

Let us start with a particular case $\Delta_i=2\Delta$. In this case, the Euclidean three point function simplifies
\be
\langle 0|  {O}( x_1)\mathcal{O}_i( x) \mathcal{O}(x_2)|0 \rangle_\mt{E}\Big|_{\Delta_i=2\Delta} = {C \over |x_1-x|^{2\Delta}  |x_2-x|^{2\Delta}}~.
\ee
Adding the appropriate small imaginary components to the times yields
\be
 \text{Re}\, \langle 0| \Big[\mathcal{O}(t_1,\vec x_1),\mathcal{O}_i(t,\vec x)\Big]\mathcal{O}(t_2, \vec x_2)|0\rangle
 \Big|_{t\geq t_1\geq t_2}
 =   {2\, C \sin^2(\pi\Delta)\over \big( - s^2_1 \big)^\Delta \big( -s^2_2 \big)^\Delta} 
  ~\Theta(-s^2_1) \Theta(-s^2_2) ~,
\ee
where we introduced $s^2_1\equiv-(t-t_1)^2+(\vec x - \vec x_1)^{2}$ and $s^2_2\equiv-(t-t_2)^2+(\vec x - \vec x_2)^{2}$.
Plugging this expression into  \reef{1ptO2} and integrating over $\vec x_1$ and $\vec x_2$ gives us the desired leading order contribution
\be
 \delta^{(2)} \langle\mathcal{O}_i(t,\vec x)\rangle\Big|_{\Delta_i=2\Delta}=
 {4\pi^{d+1}C\over \Gamma^2(\Delta)\Gamma^2\({d-2\Delta+1\over 2 }\)} \delta\lambda^2 \(\int_{-\infty}^t dt' {f(t'/\delta t)\over (t-t')^{2\Delta-d+1}}  \)^2~.
 \label{special}
 \ee
Using \reef{1pt} it can be written simply as
\be
 \delta^{(2)} \langle\mathcal{O}_i(t,\vec x)\rangle\Big|_{\Delta_i=2\Delta}
={C\over N^2} \( \delta^{(1)} \langle\mathcal{O}(t,\vec x)\rangle \)^2 ~.
\ee

Next let us consider general $\Delta_i$. In this case \reef{1ptO2} can be written as
\be 
  \delta^{(2)} \langle\mathcal{O}_i(t,\vec x)\rangle=  C \int_{-\infty}^t d t_1 \lambda( t_1) \int _{-\infty}^{t_1} dt_2 \, \lambda( t_2) 
  \Big( I_1(T_1, \overline T_2, T) - I_1(\overline T_1, \overline T_2, T) + ~\text{c. c.}  \Big) ~,
  \label{2ndO}
\ee
where the bar over $T_i^2\equiv(t_i-t-i\epsilon)^2$ for $i=1,2$ denotes complex conjugate, $T^2\equiv(t_1-t_2-i\epsilon)^2$ and we have defined\footnote{To maintain right ordering of 3 operators in \reef{1ptO2} one has to add distinct imaginary parts $i\epsilon$ and $2i\epsilon$ to Lorenzian times of $\mathcal{O}(t_1)$ and $\mathcal{O}_i(t,\vec x)$. However, this difference is not significant in this calculation, and therefore we use $i\epsilon$ instead of $2i\epsilon$.}
\be
I_1(T_1,  T_2, T)\equiv \int d^{d-1}\vec x_1 \int d^{d-1}\vec x_2 \, {1\over \Big(\vec x_1^{\,2} - T_1^{\,2}\Big)^{\Delta_i\over 2} \Big(\vec x_2^{\,2} - T_2^{\,2}\Big)^{\Delta_i\over 2} 
\Big( (\vec x_1-\vec x_2)^2 - T^2\Big)^{2\Delta - \Delta_i\over 2}}~.
\label{I_1}
\ee

To evaluate the above integral we make use of the Mellin-Barnes representation. This procedure is straightforward but tedious. The details of this calculation are presented in the Appendix \ref{masI_1}, the final answer is given by
\bea
%\scalebox{1.0}{$
&& I_1(T_1,  T_2, T)=\big(-T^2\big)^{d-1-{\Delta_i\over 2}  - \Delta} 
 {\pi^{d-1}  \over \Gamma\({d-1\over 2}\) \Gamma^2\({\Delta_i\over 2}\) \Gamma\( {2\Delta-\Delta_i\over 2}\)} 
 % $}
 \non \non
&&\times\Bigg[  \Gamma\Big({2\Delta-d+1-\Delta_i \over 2} \Big) \Gamma\Big( {d-1-\Delta_i\over 2} \Big) \Gamma\Big({2\Delta_i-d+1 \over 2} \Big)\Gamma\({\Delta_i\over 2}\)
 \non \non
&& %\scalebox{1.0}{$
 \times ~  z_2^{d-1-2\Delta_i\over2}
F_4\({\Delta_i\over 2}, {2\Delta_i - d +1\over 2};\, {3+\Delta_i-d\over 2},\, {d+1+\Delta_i-2\Delta\over 2};\, {z_1\over z_2}, {1\over z_2} \)
 %$}
 \non\non
 &&%\scalebox{1.0}{$
+  \Gamma\Big({2\Delta-d+1-\Delta_i \over 2} \Big) \Gamma\Big( {d-1-\Delta_i\over 2} \Big) \Gamma\Big({d-1 \over 2} \Big) \Gamma\({\Delta_i\over 2}\)
 \non\non
 &&\times~ 
 z_1^{d-1-2\Delta_i\over2} \({z_1\over z_2}\)^{\Delta_i\over 2}
  F_4\({\Delta_i\over 2}, {d-1\over 2};\, {d+1-\Delta_i\over 2},\, {d+1+\Delta_i-2\Delta\over 2};\, {z_1\over z_2}, {1\over z_2} \)
 %$}
\non
 \label{I1fin}
\\
 &&%\scalebox{1.0}{$
+  \Gamma\Big({2\Delta-d+1\over 2} \Big) \Gamma\Big( {d-1-\Delta_i\over 2} \Big) \Gamma\Big({d-1 +\Delta_i\over 2}-\Delta \Big) \Gamma\({\Delta_i\over 2}+\Delta-d+1\)
 \non\non
 &&\times~ 
 z_2^{d-1-\Delta- {\Delta_i\over2} } 
  F_4\({\Delta_i\over 2}+\Delta-d+1, \Delta-{d-1\over 2};\, {3-d+\Delta_i\over 2},\, {3-d-\Delta_i\over 2};\, {z_1\over z_2}, {1\over z_2} \)
 %$}
 \non\non
 && 
+  \Gamma\Big({d-1-2\Delta+\Delta_i \over 2} \Big) \Gamma\Big( {\Delta_i+1-d\over 2} \Big) \Gamma\Big(\Delta-{d-1 \over 2} \Big) \Gamma\(\Delta-{\Delta_i\over 2}\)
 \non\non
 &&\scalebox{1.1}{$\times~ 
 z_1^{d-1-{\Delta_i\over2}-\Delta} \({z_2\over z_1}\)^{{d-1\over 2} - \Delta}
  F_4\(\Delta-{\Delta_i\over 2}, \Delta-{d-1\over 2};\, {d+1-\Delta_i\over 2},\, {3-d-\Delta_i\over 2};\, {z_1\over z_2}, {1\over z_2} \) \Bigg] ~.
 $}
\non\nonumber
 \eea
Here $z_i=T_i^2/T^2$ are two dimensionless parameters and $F_4(a,b;\,c,d;\, x,y)$ is the Appell's hypergeometric function of two variables (it is defined in \reef{Appell}). 

As a consistency check we take the limit $\Delta_i\to2\Delta$. 
As expected, $I_1$ in this case dramatically simplifies. Because of the overall vanishing factor $1/\Gamma(\Delta-\Delta_i/2)$, all the terms but the last one in \reef{I1fin} vanish, and we end up with
\be
 I_1(T_1,  T_2, T)\Big|_{\Delta_i=2\Delta} = {\pi^{d-1} \Gamma^2\( \Delta-{d-1\over 2} \) \over \Gamma^2(\Delta)} \(T_1^{\,2} \, T_2^{\,2}\)^{{d-1\over 2}-\Delta} ~.
\ee 
Substituting this into \reef{2ndO} we recover \reef{special}.

The general expression for $\delta^{(2)} \langle\mathcal{O}_i(t,\vec x)\rangle$ is bulky, but it is straightforward to use it to derive the universal scaling  in the limit $\delta t\rightarrow 0$. Setting for simplicity $t=0$ yields
\be
\label{1Oscaling}
\delta^{(2)} \langle\mathcal{O}_i\rangle\Big|_{t=0} \sim (\delta t)^{-\Delta_i} \({\ell\over \delta t} \)^{2(\Delta-d)} ~.
\ee
In particular, $\langle\mathcal{O}_i\rangle$ diverges in the limit $\delta t\to 0$ if $\Delta_i> 2(d-\Delta)$. Moreover, the latter is always the case for $\mathcal{O}_i$ satisfying unitarity bound $\Delta_i>(d-2)/2$,  provided that $(3d+2)/4<\Delta<d$.

An important corollary of our calculation is the estimate of the validity of perturbation theory. When $\Delta_i=\Delta$ perturbation theory is valid so far $\delta^{(2)} \langle\mathcal{O}\rangle\Big.\ll \delta^{(1)} \langle\mathcal{O}\rangle\Big.$.
In the limit $t\gg\delta t$,  \reef{2ndO} gives
\be 
 \delta^{(2)} \langle\mathcal{O}\rangle\Big|_{t\gg\delta t} = \delta t^{-\Delta} \({\ell\over \delta t} \)^{2(\Delta-d)} 
 \bigg(  \tilde b_1\({t\over \delta t} \)^{{d-1\over 2} -\Delta}\big(1 +\ldots \big)+ \tilde b_2 \({t\over \delta t} \)^{d-1 -  {3\over 2}\Delta }\big(1 +\ldots\big) \bigg) ,
\ee
where ellipsis stand for $\mathcal{O}(\delta t/t)$ terms, and  numerical coefficients $\tilde b_i$ are of order one. 
Comparing this with  \reef{1stO} gives two conditions, with the dominant (a more restrictive) one being
\be
\label{validity}
 \left\{\begin{matrix} 
      &    t\ll\delta t \({\ell\over \delta t} \)^{2(d-\Delta)\over 2\Delta-d+1}   ~, \quad\quad~          d-1<\Delta<d ~,
      \\
      \\
      & \ \,  t\ll\delta t \({\ell\over \delta t} \)^{2(d-\Delta)\over \Delta}   ~, \quad\quad~          {d-2\over 2} <\Delta< d-1  ~. 
\end{matrix} \right.
\ee
Presence of more than one condition may indicate there are several mechanisms in place restricting the validity of the conformal perturbation theory. It is interesting to note that when the two conditions coincide, which happens for $\Delta=d-1$, the validity condition \eqref{validity} coincides with the qualitative estimate  \eqref{maxtime}.

\section{Quenched correlators}

In this section we study the response of the system to a fast quantum quench as reflected in the two-point correlation function of primary operators $\mathcal{O}_i$ and $ \mathcal{O}_j$ with the respective conformal dimensions $\Delta_i$ and $\Delta_j$.  The relevant deformation of $H_\mt{CFT}$ and its conformal dimension will be denoted by $\mathcal{O}_k$ and $\Delta_k$ respectively. 

To justify the conformal perturbation theory we require
\be
 \delta t,  t, r \ll \ell ~,
\ee
where $t$ and $r$ are the characteristic temporal and spatial separations of operators $\mathcal{O}_i$ and $ \mathcal{O}_j$. In other words, we probe the theory sufficiently close to the UV fixed point, such that the IR scale $\ell$ introduces only perturbatively small corrections within the UV CFT.

Our primary goal is to derive the universal scaling of the two-point function in various limits. 
We start by studying the case when two operators $\mathcal{O}_i$ and $ \mathcal{O}_j$ are inserted simultaneously at different points in space, and then extend our analysis to the opposite regime when both operators are inserted at the same spatial point, but at two different times. Finally, in section \ref{OPE} we show that various scalings obtained in this section can be reproduced with the help of the OPE.

At late times  and large distances $\delta t \ll t, r \ll \ell$ the equal time  correlator $ \langle\mathcal{O}_i(t, r) \mathcal{O}_j(t,0)\rangle$ after a smooth quench approaches that  one in the instantaneous quench scenario. However, at early times $t\sim \delta t$ these two scenarios disagree even if the spatial distance is large $\delta t\ll r \ll \ell$. Moreover, the disagreement also persists for late time $\delta t\ll t\ll \ell$ temporal correlator of two primaries, $ \langle\mathcal{O}_i(t,0) \mathcal{O}_j(0)\rangle$. This confirms the expectation of \cite{Das:2014jna,Das:2014hqa,Das:2015jka} that these two protocols result in two very different states after the quench, as explained in section \ref{WF}.

\subsection{Spatial correlators}

In the case of spatially separated operators the analog of \reef{1ptO} takes the following form
\bea
&& \langle \mathcal{O}_i(t,\vec x)\mathcal{O}_j(t,0)\rangle =  \langle 0| \mathcal{O}_i(t,\vec x)\mathcal{O}_j(t,0)|0\rangle
\label{corr}
 \\
 &&- i \int_{-\infty}^t dt' \, \lambda(t') \int d^{d-1}\vec y \, 
   \langle 0| [\mathcal{O}_i(t,\vec x)\mathcal{O}_j(t, 0) , \mathcal{O}_k(t', \vec y) ] |0\rangle + \mathcal{O}(\delta\lambda^2) ~.
 \nonumber
\eea
Of course, $\langle 0| \mathcal{O}_i\mathcal{O}_j|0\rangle=0$ unless $\Delta_i=\Delta_j$.  The three point function in the integrand is obtained by an appropriate analytic continuation of its Euclidean counterpart. For simplicity we introduce 
\beq
 \Delta_{ijk}\equiv\Delta_i+\Delta_j-\Delta_k ~.
\eeq
Then the Euclidean three point function reads
\be
\langle 0|  {O}_i(x)\mathcal{O}_j(y) \mathcal{O}_k(z)|0 \rangle_\mt{E} = {C_{ijk} \over |x-y|^{\Delta_{ijk}} |x-z|^{\Delta_{kij}} |y-z|^{\Delta_{kji}}}~.
\label{Eucl3p}
\ee
The Lorentzian ordered correlator that we need can be obtained from the above Euclidean expression by adding small imaginary component to the Lorentzian times of each operator\footnote{See Appendix \ref{intcom} for a simple example.} 
\bea
&& \langle 0| [\mathcal{O}_i(t,\vec x)\mathcal{O}_j(t,0), \mathcal{O}_k(0, \vec y) ] |0\rangle = 
\non
&&{C_{ijk}\over \big( -(t-i\epsilon)^2 + \vec y^{\,2} \big)^{\Delta_{kji}\over 2}\big( -(t-i\epsilon)^2 + (\vec y-\vec x)^2 \big)^{\Delta_{kij}\over 2}
|\vec x|^{\Delta_{ijk}} } 
\label{comm3O} \\
&& -{C_{ijk}\over \big( -(t+i\epsilon)^2 + \vec y^{\,2} \big)^{\Delta_{kji}\over 2}\big( -(t+i\epsilon)^2 + (\vec y-\vec x)^2 \big)^{\Delta_{kij}\over 2}
|\vec x|^{\Delta_{ijk}} } ~.
\nonumber
\eea
In principle, one can repeat now the same steps as in Appendix \ref{intcom} and get the desired formulas. However, the calculations become a bit cluttered because of proliferation of theta functions (see Appendix \ref{noMB}). Hence, we do it in a slightly different way using Mellin-Barnes representation \reef{MB}.  

From \reef{corr}  linear response to the quench (leading correction to a pure CFT two-point function) can be written as
\be
 \delta^{(1)}\langle \mathcal{O}_i(t,\vec x)\mathcal{O}_j(t,0)\rangle = {2 C_{ijk}\over x^{\Delta_{ijk}} } 
 \int_{-\infty}^t dt' \, \lambda(t') ~ \text{Im}\big( \,J(t-t',x) \big)~,
 \label{OOspace}
\ee
where we have defined
\be
 J(t,x)\equiv
 \int d^{d-1}\vec y \, {1\over \big( -(t-i\epsilon)^2 + \vec y^{\,2} \big)^{\Delta_{kji}\over 2}\big( -(t-i\epsilon)^2 + (\vec y-\vec x)^2 \big)^{\Delta_{kij}\over 2} } ~.
\ee
Introducing Feynman parameter $u$ to integrate over $\vec y$, yields
\be
 J(t,x)={ \pi^{d-1\over 2} \Gamma\( \Delta_k-{d-1\over 2} \) 
 \over \Gamma\big( {\Delta_{kji}\over 2} \big) \Gamma\big( {\Delta_{kij}\over 2} \big)}
 \int_0^1 u^{{\Delta_{kij}\over 2}-1} (1-u)^{{\Delta_{kji}\over 2}-1} \big(u(1-u)x^{\,2}-(t-i\epsilon)^2\big)^{{d-1\over 2}-\Delta_k}~.
\ee
To carry out integration over the Feynman parameter $u$ we employ Mellin-Barnes representation \reef{MB} with $\nu=\Delta_k-{d-1\over 2}$, $A^2=u(1-u)x^2$ and $M^2=(t-i\epsilon)^2$,
\bea
&&J(t,x)={ \pi^{d-1\over 2}   \over 2\pi i \,  \Gamma\big({\Delta_{kji}\over 2} \big) \Gamma\big( {\Delta_{kij}\over 2} \big)} ~
 \\
 &&\times \int_{c-i\infty}^{c+i\infty} ds ~  {\big(-(t-i\epsilon)^2\big)^s\over (x^{\,2})^{\nu+s}} 
  { \Gamma(-s) ~ \Gamma(\nu+s) \Gamma\({\Delta_{kij}\over 2}-\nu-s\) \Gamma\({\Delta_{kji}\over 2}-\nu-s\) \over \Gamma\({d-1\over 2}-\nu-2s\)} ~.
  \nonumber
\eea
For $|M^2/A^2|>1$ we close the contour to the left encompassing the infinite series of poles of $\Gamma(\nu+s)$ and possibly finite number of poles associated with $\Gamma\({\Delta_{kij}\over 2}-\nu-s\)$ and  $\Gamma\({\Delta_{kji}\over 2}-\nu-s\)$. However, recall that we analytically continue various parameters (such as $d,\nu$ and scaling dimensions) to the values where the integrals converge. Other values are treated by analytic continuation. In particular, both $\Delta_{kij}$ and $\Delta_{kji}$ are positive to ensure convergence of the integral over Feynman parameter $u$. Thus the poles of $\Gamma(\nu+s)$ are separated from the poles of other gamma functions occurring in the above integral, and we can readily evaluate the sum over the residues of $\Gamma(\nu+s)$. The final result is given by
\be
\small
J(t,x)={\pi^{d-1\over 2} \Gamma\(\Delta_k-{d-1\over 2}\) \over \Gamma(\Delta_k)\big(-(t-i\epsilon)^2\big)^{\Delta_k-{d-1\over 2}} }   ~
{}_3F_2  \({\Delta_{kij}\over 2}, \, {\Delta_{kji}\over 2}, \, \Delta_k-{d-1\over 2} \,; {\Delta_k\over 2}, \, {\Delta_k+1\over 2} \, ; \, {x^2\over 4 (t-i\epsilon)^2} \).
\ee
The bounds on various parameters which we imposed to ensure convergence of the integrals can be relaxed now. Substituting the above result back into \reef{OOspace} yields
\bea
&& \delta^{(1)}\langle \mathcal{O}_i(t,\vec x)\mathcal{O}_j(t,0)\rangle = { 2\,\pi^{d-1\over 2}  \Gamma\(\Delta_k - {d-1\over 2}\) \over  \Gamma(\Delta_k)  } 
~ {C_{ijk} \over x^{\Delta_{ijk}}} ~ \text{Im} \int_{-\infty}^t dt' \, {\lambda(t')\over \big(-(t-t'-i\epsilon)^2\big)^{\Delta_k - {d-1\over 2}}} 
\non
&&\quad\quad\quad\quad\quad\times  
 ~ {}_3F_2  \({\Delta_{kij}\over 2}, \, {\Delta_{kji}\over 2}, \,\Delta_k-{d-1\over 2} \,; {\Delta_k\over 2}, \, {\Delta_k+1\over 2} \, ; \, {x^2\over 4 (t-t'-i\epsilon)^2} \) .
 \label{OOsfin}
\eea
Obviously, one can suppress $i\epsilon$ in the argument of the generalized hypergeometric function in the region where it is analytic. 

Equation \reef{OOsfin} is convenient to explore various limits. For instance, the late time behavior of the linear response is given by 
\be
\delta^{(1)}\langle \mathcal{O}_i(t,\vec x)\mathcal{O}_j(t,0)\rangle\Big|_{x, \delta t\, \ll t} \simeq {- 2\pi^{d+1\over 2} C_{ijk} \over \Gamma(\Delta_k)\Gamma\({d+1 - 2\Delta_k\over 2}  \)}
 { \delta t  \, \delta\lambda \, t^{d-1-2\Delta_k}\over x^{\Delta_{ijk}}}\int_{-\infty}^{\infty} d\xi f(\xi)     ~.
 \label{2plarget}
\ee 
In the limit of fast and smooth quenches $\delta t\to 0$ this contribution vanished, which  agrees with the behavior in case of a sudden quench. Of course, the genuine late time behavior $t\gg \ell$ requires a more elaborate analysis since the conformal perturbation theory is not reliable in this regime. We also remark, that when 
$\Delta_k={d-1\over 2}$, as is the case for free fermion mass operator, the two-point function becomes $t$-independent. 

Next we turn to study early times when the conformal perturbation theory is valid. Setting $t=0$ for simplicity and considering $x\gg\delta t$ and $x\ll \delta t$ gives
\bea
 \delta^{(1)}\langle \mathcal{O}_i(0,\vec x)\mathcal{O}_j(0,0)\rangle\Big|_{ x\gg \delta t}
&\simeq& {- 2 \pi^{d+1\over 2} \over \Gamma\big( {\Delta_{kji}\over 2} \big)\Gamma\big({d-\Delta_{kji}+1\over 2}\big)} ~{C_{ijk} \over  x^{2\Delta_i}}   
{\delta\lambda \over (\delta t)^{\Delta_{kji}-d} }
 \int_{-\infty}^0 d\xi {f(\xi) \over (-\xi )^{\Delta_{kji}-d+1}}
   \non
   \non
  &+&(i\leftrightarrow j)~ , 
\label{2pscaling}
\\
 \delta^{(1)}\langle \mathcal{O}_i(0,\vec x)\mathcal{O}_j(0,0)\rangle\Big|_{ x\ll\delta t} &\simeq& {- 2\pi^{d+1\over 2}  \over  \Gamma\( \Delta_k \)  \Gamma\({d-2\Delta_k+1\over 2}\)  } 
~ {C_{ijk} \over x^{\Delta_{ijk}}} {\delta\lambda\over (\delta t)^{2\Delta_k-d}} \int_{-\infty}^0 d\xi \, {f(\xi)\over (-\xi)^{2\Delta_k - d+1}} ~.
 \nonumber
\eea
Note that the integrals over $\xi$ are finite if we employ dimensional regularization scheme and choose the scaling dimensions such that the logarithmic divergence is not present. 

As can be seen from \reef{2pscaling} the two-point function becomes singular in the limit of fast quenches $\delta t\to 0$ while $\delta\lambda$ and $x$ are held fixed, if $\Delta_{kij}> d$ or $\Delta_{kji}> d$. Our calculation demonstrates that during the quench the universal scaling of spatial correlators flows from $\delta t^{d-2\Delta_k}$ when $x\sim \delta t$ to $\delta t^{d-\Delta_{kji}}$ when $x\gg \delta t$. 

Finally, we note that for the special conformal dimensions when  $n=\Delta_{kji}-d+1$ is integer and odd there is a logarithmic enhancement of the scaling in \reef{2pscaling}, while for even integer $n$ this scaling is balanced by zero of the gamma function in the denominator of \reef{2pscaling}.\footnote{Similar argument holds for two other terms in \reef{2pscaling} with $2\Delta_k$ and $\Delta_{kij}$ playing the role of $\Delta_{kji}$.} This behavior is similar to the one discussed above \reef{fin1pt}. In particular, for  $t\sim \delta t$ we find
\bea
%\label{2ptspatial}
  &&\delta^{(1)}\langle \mathcal{O}_i(t,\vec x)\mathcal{O}_j(t,0)\rangle\Big|_{ x\gg \delta t,\,\Delta_{kji}=d+n-1}=
   {  2\pi^{d-1\over 2}  \Gamma\big({n\over 2}\big) \over  \Gamma\big({d-1+n\over 2}\big)  \Gamma(n)} {C_{ijk} \over |\vec x|^{2\Delta_i}} {d^{n-1}\over dt^{n-1}}\lambda(t)
   \nonumber
   \\
 &&\quad\quad\times \left\{\begin{matrix} 
      &   (-1)^{n\over 2}  {\pi \over  2}  \quad\quad\quad\quad\quad~\, , \, \text{for {\bf even}} ~ n\in\mathbb{N}^+ ~,
      \\
      \\
      &   (-1)^{n+1\over 2} \log\big(a_0/ \delta t\big) \quad , \, \text{for {\bf odd} } ~ n\in\mathbb{N}^+ ~.
 \end{matrix} \right.
\eea

\subsection{Temporal correlators}

Now let us study the case when $\mathcal{O}_i$ and $ \mathcal{O}_j$ are inserted at the same spatial point, but at two different times. Using (\ref{state}) we find the following linear response
\bea
&& \delta^{(1)} \langle \mathcal{O}_i(t_1,0)\mathcal{O}_j(t_2,0)\rangle =  %\langle \mathcal{O}_i(t_1, 0)\mathcal{O}_j(t_2,0)\rangle_{CFT} 
 - i \int_{-\infty}^{t_2} dt' \int d^{d-1}\vec y \, 
 \lambda(t') \langle 0| \big[\mathcal{O}_i(t_1,0)\mathcal{O}_j(t_2, 0) , \mathcal{O}_k(t', \vec y) \big] |0\rangle 
\nonumber
  \\
 &&- i \int_{t_2}^{t_1} dt' \int d^{d-1}\vec y \, 
 \lambda(t') \langle 0| \big[\mathcal{O}_i(t_1,0) , \mathcal{O}_k(t', \vec y) \big] \mathcal{O}_j(t_2, 0) |0\rangle ~. %+ \mathcal{O}(\delta\lambda^2) ~,
 \label{tempcorr}
\eea
Then using   \reef{Eucl3p}  we have 
\bea
&& \langle 0| [\mathcal{O}_i(t_1,0)\mathcal{O}_j(t_2,0), \mathcal{O}_k(0, \vec y) ] |0\rangle = 
\\
&&{C_{ijk}\over \big( -(t_2-i\epsilon)^2 + \vec y^{\,2} \big)^{\Delta_{kji}\over 2}\big( -(t_1-2i\epsilon)^2 + \vec y^{\,2} \big)^{\Delta_{kij}\over 2}
\big(-(t_1-t_2-i\epsilon)^2\big)^{\Delta_{ijk}\over 2} } 
\nonumber \\
&& -{C_{ijk}\over \big( -(t_2+2i\epsilon)^2 + \vec y^{\,2} \big)^{\Delta_{kji}\over 2}\big( -(t_1+i\epsilon)^2 + \vec y^{\,2} \big)^{\Delta_{kij}\over 2}
\big(-(t_1-t_2-i\epsilon)^2\big)^{\Delta_{ijk}\over 2}  } ~,
\nonumber
\eea
and 
\bea
&& \langle 0| \big[\mathcal{O}_i(t_1,0) , \mathcal{O}_k(0, \vec y) \big] \mathcal{O}_j(t_2, 0) |0\rangle = 
\\
&&{C_{ijk}\over \big( -(t_2+i\epsilon)^2 + \vec y^{\,2} \big)^{\Delta_{kji}\over 2}\big( -(t_1-i\epsilon)^2 + \vec y^{\,2} \big)^{\Delta_{kij}\over 2}
\big(-(t_1-t_2-2i\epsilon)^2\big)^{\Delta_{ijk}\over 2} } 
\nonumber \\
&& -{C_{ijk}\over \big( -(t_2+2i\epsilon)^2 + \vec y^{\,2} \big)^{\Delta_{kji}\over 2}\big( -(t_1+i\epsilon)^2 + \vec y^{\,2} \big)^{\Delta_{kij}\over 2}
\big(-(t_1-t_2-i\epsilon)^2\big)^{\Delta_{ijk}\over 2}  } ~.
\nonumber
\eea
To maintain right ordering of various operators in \reef{tempcorr} we added two small imaginary parts $i\epsilon$ and $2i\epsilon$ to the appropriate Lorentzian times. However, for the calculations we carry out in th section this difference between  $i\epsilon$ and $2i\epsilon$ matters. 

Substituting these expressions into \reef{tempcorr} and integrating over $\vec y$, gives
\bea
 && \delta^{(1)} \langle \mathcal{O}_i(t_1,0)\mathcal{O}_j(t_2,0)\rangle = {4\pi^{d-1\over 2}\, C_{ijk}\over\Gamma\({d-1\over 2}\)\big(-(t_1-t_2-i\epsilon)^2\big)^{\Delta_{ijk}\over 2}} \int_{-\infty}^{t_2} dt' \, 
 \lambda(t')  \text{Im}\big(I_2(T_1,\,T_2)\big)
  \non
 && -i \, {2\pi^{d-1\over 2}\, C_{ijk} \over\Gamma\({d-1\over 2}\)\big(-(t_1-t_2-i\epsilon)^2\big)^{\Delta_{ijk}\over 2}} \int_{t_2}^{t_1} dt' \, 
 \lambda(t') \( I_2(T_1,\,\overline T_2) - I_2( \overline T_1,\,\overline T_2) \) ~,
 \label{tempOO}
\eea
where the bar over $T_i^2\equiv(t_i-t'-i\epsilon)^2$ denotes complex conjugate, and we have defined 
\bea
&&I_2(T_1,\,T_2)\equiv\int d y\scalebox{0.97}{$
{y^{d-2}\over \big( -T_1^{\,2} +  y^{\,2} \big)^{\Delta_{kij}\over 2} \big( -T_2^{\,2} +  y^{\,2} \big)^{\Delta_{kji}\over 2} } 
$}
\label{I}
\\
\non
&&\scalebox{1.05}{$
~=\big(-T_2^{\,2}\big)^{d-1-2\Delta_k\over 2} \( { \Gamma\({d-1\over 2}\) \Gamma\({\Delta_{kji}-d+1\over 2}\)\over 2\Gamma\({\Delta_{kji}\over 2}\) } 
\({-T_2^{\,2}\over -T_1^{\,2}}\)^{{\Delta_{kij}\over2}}
 {}_2F_1  \({d-1\over 2},\,{\Delta_{kij}\over 2}, {d+1-\Delta_{kji}\over 2}; \, {T_2^{\,2}\over  T_1^{\,2}}\) \right.
 $}
 \non
 \non
&&\scalebox{1.05}{$\quad\quad \left. +
 { \Gamma\({d-1-\Delta_{kji}\over 2}\) \Gamma\({1-d+2\Delta_k\over 2}\)\over 2\Gamma\({\Delta_{kij}\over 2}\) } \({-T_2^{\,2}\over -T_1^{\,2}}\)^{1+2\Delta_{k}-d\over2}
 {}_2F_1  \({\Delta_{kji}\over 2}, \, {1-d+2\Delta_{k}\over 2},\, {3-d+\Delta_{kji}\over 2}; \, {T_2^{\,2}\over  T_1^{\,2}}\) \) . $}
 \nonumber 
 \eea

Equation \reef{tempOO} combined with \reef{I} is what we need to analyze various limits. For instance, if $t_1,t_2\gg \delta t$ then it can be readily seen that the linear response of the temporal correlator vanishes as $\delta t\to 0$.  However, an interesting scaling emerges in the limit of fast but smooth quenches if, for example, $t_1\gg\delta t$ while $t_2$ is set at some value within the support of $\lambda(t)$. 
Indeed, setting for simplicity $t_2=0$ and assuming sufficiently large $\Delta_j$, one gets in this regime
\bea
\label{420formula}
&&%\scalebox{0.97}{$
I_2(T_1,\,T_2)\Big|_{t_1\gg\delta t, t_2=0}\simeq \big(-T_2^{\,2}\big)^{d-1-2\Delta_k\over 2}  { \Gamma\({d-1\over 2}\) \Gamma\({\Delta_{kji}-d+1\over 2}\)\over 2\Gamma\({\Delta_{kji}\over 2}\) } 
\({-T_2^{\,2}\over -T_1^{\,2}}\)^{{\Delta_{kij}\over2}} ~.
% $}
 \nonumber 
 \eea
Substituting into \reef{tempOO}, yields
\bea
\label{2pttime}
 && \delta^{(1)} \langle \mathcal{O}_i(t_1,0)\mathcal{O}_j(0,0)\rangle\Big|_{t_1\gg\delta t}
  \simeq 
  \\
  \non
 &&+ {2\pi^{d-1\over 2}\, \Gamma\({\Delta_{kji}-d+1\over 2}\) \sin\big({\pi(d-1-2\Delta_k)\over 2}\big)\over \Gamma\({\Delta_{kji}\over 2}\) }
e^{-i{\pi\Delta_{ijk}\over 2}}  {(\delta t)^{d-\Delta_{kji}}\over t_1^{2\Delta_i}} C_{ijk} 
  \int_{-\infty}^{0} dt' \, {\lambda(t')\over (-t')^{\Delta_{kji}+1-d}}  
  \non
 && - \, {2\pi^{d-1\over 2}\, \Gamma\({\Delta_{kji}-d+1\over 2}\)\,\sin\({\pi\over 2}\Delta_{kij}\)  \over \Gamma\({\Delta_{kji}\over 2}\) }   \, e^{i{\pi(d-1-2\Delta_j)\over 2}} {(\delta t)^{d-\Delta_{kji}}\over t_1^{2\Delta_i}} C_{ijk} 
 \int_{0}^{\infty} dt' \, 
 {\lambda(t')\over t'^{\Delta_{kji}+1-d}}  ~.
 \nonumber
\eea
This expression clearly demonstrates that the two-point temporal correlator for large but fixed $t_1$ and small $t_2$  is amplified (and in fact diverges) in the limit $\delta t\to 0$ for sufficiently large $\Delta_j$.

\subsection{Universal scaling via OPE}
\label{OPE}

In this subsection we illustrate that the universal scaling \reef{2pscaling} of the quenched spatial correlator, which emerges in the limit of fast and smooth quenches,  can be recovered using the OPE. We start from the simplest regime $x\ll \delta t$. In this limit we replace  
\be
 \mathcal{O}_j(0,\vec x) \, \mathcal{O}_j(0,0) \sim {C_{ijk}\over N_i  x^{\Delta_{ijk}} } \, \mathcal{O}_i(0) + \ldots~,
\ee
where $N_i$ is normalization constant of the Euclidean correlator $\langle\mathcal{O}_i\mathcal{O}_i\rangle_\mt{E}$. As a result, the problem of computing \reef{corr} reduces to \reef{1ptO} where $\Delta$ is identified with $\Delta_k$. In particular, combining \reef{1pt} with the coefficient of the above OPE yields the desired formula appearing in the second line of \reef{2pscaling}. Similarly, one can derive late time behavior \reef{2plarget}.

To understand how to use OPE in the limit $x\gg \delta t$, we rewrite the string of operators appearing in the integrand of \reef{corr} as follows
\be
\nonumber
  [\mathcal{O}_i(0,\vec x)\mathcal{O}_j(0, 0) , \mathcal{O}_k(t', \vec y) ] = 
  \mathcal{O}_i(0,\vec x) [\mathcal{O}_j(0, 0) , \mathcal{O}_k(t', \vec y) ]  +  [\mathcal{O}_i(0,\vec x) , \mathcal{O}_k(t', \vec y) ] \mathcal{O}_j(0, 0) ~.
\ee 
Both commutators on the right hand side vanish unless $\mathcal{O}_k$ sits  inside the past/future light cone centered at the insertion points of $\mathcal{O}_i$ or $\mathcal{O}_j$.  Now since $\lambda(t)$ vanishes outside the time interval of order $\delta t$, we deduce that in the regime $x\gg\delta t$ the relevant domains of two light cones are disjoint small neighborhoods of $\mathcal{O}_i$ and $\mathcal{O}_j$ respectively. Their size is of order $\delta t$ and they  are separated by a large space-like distance of order $|\vec x|$. Thus for $O_k$ sitting within these domains, only one of the commutators on the right hand side survives, and OPE can be used to replace it. 
For example, in Euclidean space we have
\be
 \mathcal{O}_j(0,0) \, \mathcal{O}_k(t_\mt{E}', \vec y) \sim {C_{ijk}\over N_i(t_\mt{E}^{'\,2}+\vec y^{\, 2})^{\Delta_{jki}\over 2}} \, \mathcal{O}_i(0) + \ldots~.
 \label{commut}
\ee
Hence, following the $i\epsilon$ prescription outlined in the previous section, we obtain
\be
 [\mathcal{O}_j(0, 0) , \mathcal{O}_k(t', \vec y) ]  \sim 2\, i \, {C_{ijk}\over N_i}\, \mathcal{O}_i(0,0) \,   \text{Im} \,{1\over (-(t'+i\epsilon)^2+\vec y^{\, 2})^{\Delta_{kji}\over 2}}  + \ldots~.
\ee
or equivalenly,
\be 
  [\mathcal{O}_j(0, 0) , \mathcal{O}_k(t', \vec y) ]  \sim 2\, i \, {C_{ijk}\over N_i} \, \mathcal{O}_i(0,0) \, 
  {\Theta(-s^2)\over (-s^2)^{\Delta_{kji}\over 2}} \sin\( {\pi\Delta_{kji}\over 2} \) \text{sign}(t') ~,
\ee
where $s^2=-t^{'\,2}+\vec y^{\, 2}$. Of course, up to the trivial replacements $i\leftrightarrow j$ and $\vec y\to \vec y - \vec x$, this relation also holds for $[\mathcal{O}_i(0, \vec x) , \mathcal{O}_k(t', \vec y) ]$. Pluging this expression back into \reef{commut} and \reef{corr}, yields
\be
 \delta^{(1)}\langle \mathcal{O}_i(0,\vec x)\mathcal{O}_j(0,0)\rangle\Big|_{|\vec x| \gg\delta t} = 
 - 2\,C_{ijk} {\sin\big( {\pi\Delta_{kji}\over 2} \big) \over |\vec x|^{2\Delta_i}}  \int_{-\infty}^0 dt' \lambda(t') \int d^{d-1}y {\Theta(-s^2)\over (-s^2)^{\Delta_{kji}\over 2}} 
 +(i\leftrightarrow j)~ .
\ee
The integral over $\vec y$ is straightforward, and the final answer matches the first expression in \reef{2pscaling}.

\section{Conclusions}
In this paper we discussed global quantum quenches in field theory, focusing on the regime when the typical time-scale of interaction $\delta t$ is much shorter than all other physical scales except for the UV cutoff. We outlined qualitative time evolution of the wave-function following the quench and argued that for the times not exceeding critical value \eqref{maxtime} behavior of the system can be described in terms of the conformal perturbation theory. Finally, we used conformal perturbation theory to calculate time-dependence of one- and two-point correlation functions of scalar primaries of {\it arbitrary} dimensions and established new universal scaling behavior for these quantities: \reef{1stO}, \reef{1Oscaling},\reef{2pscaling},\reef{2pttime}.

Our results raise a number of interesting questions. Besides time evolution of local quantities, which were studied in this paper, it would be interesting to use conformal perturbation theory to evaluate dynamics of non-local quantities as well, e.g.~growth and spread of entanglement entropy following the quench \cite{David:2016pzn,Caputa:2017ixa,David:2017eno}. Of particular interest would be to shed light on universal behavior of entanglement which was previously established holographically in \cite{Liu:2013iza,Liu:2013qca}.

One of the results of this paper is the prediction of thermalization, that following the fast global quench local observables eventually equilibrate to their respective thermal expectation values. Dynamics of thermalization goes beyond the scope of the conformal perturbation theory, but still should be described in terms of the non-perturbative CFT dynamics. This gives hope that relaxation dynamics may exhibit some universal scaling behavior.  Conceivably, such a universal scaling can be established numerically in case of (1+1) dimensional models \cite{Das:2017sgp}, see also \cite{Delfino:2016bln} for numerical studies of instantaneous global quenches near criticality. We hope to address this and other related questions in the future.

\acknowledgments  
We thank Sumit Das, Michael Eides, Shmuel Elitzur, Damian Galante, Barak Kol, Eliezer Rabinovici and Ruth Shir for helpful discussions.  
This work is supported by the BSF grant 2016186. The research of MS is supported by the "Quantum Universe" I-CORE program of the Israel Planning and Budgeting Committee (grant 1937/12) and partially by a grant from the Simons Foundation. The work was done in part at Aspen Center for Physics, which is supported by National Science Foundation grant PHY-1607761.

\appendix

\section{Transition probability: direct calculation}
\label{trans}

In what follows we derive the total probability for vacuum decay using the traditional technique of time-dependent perturbation theory. If we perturb the Hamiltonian $H\rightarrow H+ \mathcal{V}(t)$, then the expression for the first order transition amplitude to an eigenstate $|n\rangle$ of the unperturbed Hamiltonian reads \cite{LL}
\beq
 a_{n0}=-i\int_{-\infty}^t \langle 0|\mathcal{V}(t')|n\rangle e^{iE_n t'} \, dt' = - {e^{iE_n t}\over E_n} \langle 0|\mathcal{V}(t)|n\rangle + {1\over E_n} \int_{-\infty}^t \langle 0|{\partial \mathcal{V}(t') \over \partial t'}|n\rangle e^{iE_n t'}  ~.
 \label{amp}
\eeq
The first term on the right hand side is the first order correction to the ground state wave function due to the perturbation. It has nothing to do with transition amplitude and we suppress it in what follows. Of course, this term vanishes for sufficiently large $t$ if perturbation asymptotes to  zero. 
The decay probability $P$ to second order in $\lambda(t)$ is given by 
\bea
P&=&\sum_{n\neq 0} |a_n|^2
\\
&=& (2\pi)^{d-1} V \sum_{n\neq 0}{\delta(\vec p_n) \left|\langle 0|\mathcal{O}(0)|n\rangle\right|^2\over E_n^2} \int_{-\infty}^t dt'' e^{-iE_n t''}\dot \lambda(t'') \int_{-\infty}^t dt' e^{iE_n t'}\dot\lambda(t')   + \ldots\ ,
\nonumber
\eea
where $V$ is the volume of space \reef{vol}, dot denotes derivative with respect to time, and we substituted $\mathcal{V}(t)= \lambda(t) \int \mathcal{O}$ into \eqref{amp} and used the identity
\bea
\langle 0|{\partial \mathcal{V}(t) \over \partial t} |n\rangle&=&\dot\lambda(t) \langle 0|\mathcal{O}(0)|n\rangle\, 
(2\pi)^{d-1}\delta^{d-1}(\vec{p}_n)\ .
\eea

If the instant $t$ is taken after the quench is over (when $\lambda(t)$ is constant), we can substitute $t\to\infty$ and rewrite decay probability in terms of the Fourier components of $\dot\lambda(t)$ as follows\footnote{ Note that dot in the case of $\dot{\lambda}(E_n)$ does not stand for the derivative with respect to time.This is just a  Fourier transform of $\dot\lambda(t)=d\lambda/dt(t)$. }
\bea
P= (2\pi)^{d-1} V \sum_{n\neq 0}{\delta(\vec p_n) \left|\langle 0|\mathcal{O}(0)|n\rangle\right|^2\over E_n^2}
|\dot{ \lambda}(E_n)|^2 +\ldots\ ,
\eea
The sum over $n$ can be carried out using the definition \eqref{rhodef}. Indeed, for any function $f(p)$ we have
\bea
\nonumber
&&\sum_n \left|\langle 0|\mathcal{O}(0)|n\rangle\right|^2 f(p_n)=\\ &&\int d^d p \sum_n f(p) \delta^d(p-p_n)  \left|\langle 0|\mathcal{O}(0)|n\rangle\right|^2=\int {d^d p\over (2\pi)^{d-1}} f(p) \rho(p^2) \Theta(p^0)\ .
\eea
Thus we get\footnote{Dashed integral stands to emphasize that contribution of the vacuum state should be excluded.}
\bea
P=  V \dashint_{0}^\infty d\omega\, \rho(\omega^2) { |\dot{ \lambda}(\omega)|^2\over \omega^2}  +\ldots\ .
 \label{probab}
\eea

If $\lambda(t)$ approaches zero in the asymptotic future, then one can replace $\dot{\tilde\lambda}(\omega)\to i \omega\tilde\lambda(\omega)$, which is the standard identity for the Fourier transform in this case. Hence, we recover transition probability used in the main body of the text
\bea
P=  V \dashint_{0}^\infty d\omega\, \rho(\omega^2) | \lambda(\omega)|^2+\ldots\ .
\eea

\section{Conformal spectral function}
\label{cftspec}

In this Appendix we calculate the conformal spectral function used in the text. Let us consider a correlation function of two scalar primaries in a $d$-dimensional Euclidean CFT, 
\bea
\langle \mathcal{O}(x) \mathcal{O}(0)\rangle ={N\over |x|^{2\Delta}}={N(4\pi)^{d/2}\Gamma(d/2-\Delta)\over 4^{\Delta} \Gamma(\Delta)}\int {d^dp\over (2\pi)^d} {e^{ipx}\over (p^2)^{d/2-\Delta}} .
\eea
The same correlation function can be rewritten in terms of K$\ddot{\rm a}$ll$\acute{\rm e}$n-Lehmann representation 
\bea
\label{KL}
{N(4\pi)^{d/2}\Gamma(d/2-\Delta)\over 4^{\Delta} \Gamma(\Delta)} {1\over (p^2)^{d/2-\Delta}}  =  \int_0^\infty \rho(\mu^2) {d\mu^2 \over p^2+\mu^2} \ ,
\eea
where the spectral function $\rho(\mu^2)$ is defined in \eqref{rhodef}
and the sum runs over the complete set of eigenstates $|n\rangle$ of the momentum operators $P^\mu$. 

Since the theory is conformal, we deduce that the spectral function is homogeneous, $\rho(\mu^2)=C\mu^{2\alpha}$. Using \eqref{KL} leads to
\bea
{N(4\pi)^{d/2}\Gamma(d/2-\Delta)\over 4^{\Delta} \Gamma(\Delta)} {1\over (p^2)^{d/2-\Delta}}  = C \, p^{2\alpha} \int_0^\infty  {x^{2\alpha} dx^2 \over 1+x^2} \ .
\eea
Integrating over $x$, yields
\beq
 \alpha=\Delta - {d\over 2}\, , \quad  C=  {N(4\pi)^{d/2}\over 4^{\Delta} \Gamma(\Delta) \Gamma\big(1-  {d\over 2} +\Delta\big)} ~.
 \eeq
Note that by definition the spectral function is positive definite, hence $\Delta\ge(d-2)/2$ to ensure positivity of $C$, which recovers the well-known unitarity bound.\footnote{When $\Delta={d-2 \over 2}$, the coefficient $C$ vanishes and according to \eqref{KL} the spectral function takes the form $\rho(\mu^2)=\tilde C \delta(\mu^2)$ with $\tilde C = {8 \pi ^{d/2} N\over  \Gamma\big({d-2 \over 2}\big)} $.}

\section{Work done on the system}
\label{Ward}

In this Appendix we derive expression for the energy density after the quench in case of a conformal theory \reef{Eden} using the Ward identity
\beq
\label{WardId}
 {d \mathcal{E} \over dt} = \partial_t\lambda(t) \langle \mathcal{O} (t,0) \rangle  ~.
\eeq
This approach was previously used  in the context of holographic and free field theory calculations  in \cite{Buchel:2012gw,Buchel:2013lla,Buchel:2013gba,Das:2014jna,Das:2014hqa,Das:2015jka}.

Note that the right hand side of the Ward identity is finite in the limit $a_0\to 0$ since we implicitly assume that $\mathcal{O}$ is renormalized and the action is equipped with all necessary counterterms to subtract the UV divergences of $\langle\mathcal{O}\rangle$. In fact, we employ analytic continuation in $d$ (or $\Delta$), and therefore only logarithmic terms in \reef{fin1pt} survive. However, as we argue below these terms do not contribute to the total energy pumped into the system during the quench. This is consistent with our previous analysis of the energy density in section \ref{WF}.

Substituting \reef{1pt} into \eqref{WardId} yields
\beq
 {d \mathcal{E} \over dt} =  {-2 \pi^{d+1\over 2} N \over  \Gamma(\Delta) \Gamma\big( {d-2\Delta+1\over 2} \big)} \partial_t\lambda(t) \int_{-\infty}^t dt'
 \,{\lambda(t') \over |t-t'|^{2\Delta -d +1}  } + \mathcal{O}(\lambda^3) ~.
\eeq
The work done on the system can be obtained by integrating over time and taking the limit $t\gg \delta t$. The leading order correction reads
\beq
 \mathcal{E}^{(2)} 
 = {- 2 \pi^{d+1\over 2} N \over  \Gamma(\Delta) \Gamma\big( {d-2\Delta+1\over 2} \big)} \, {\delta\lambda^2\over\delta t^{2\Delta-d}} \int_{-\infty}^\infty d\xi''  \int_{-\infty}^{\xi''} d\xi'
 \,{ \partial_{\xi''} f(\xi'') f(\xi') \over |\xi''-\xi'|^{2\Delta -d +1}  } ~.
 \label{work}
\eeq
The integral on the right hand side is finite, and there is no logarithmic enhancement of the scaling $\delta t^{2\Delta-d}$. Indeed, the power law divergences are irrelevant since we employ analytic continuation in $d$ or $\Delta$, while from \reef{fin1pt} all possible logarithmic terms in the integrand are proportional to $\del_t f \del_t^{n-1}f$ for some odd $n=1,2,..,d+1$. Thus up to vanishing boundary terms  $\del_t f \del_t^{n-1}f \sim (-1)^{n-1\over 2} {1\over 2} \del_t(\del_t^{n-1\over 2}  f)^2$  is a total derivative which vanishes upon integration. Finally, integrating \reef{work} by parts and using Fourier representation of $\lambda(t)$ results in 
\bea
 \mathcal{E}^{(2)} =
 {(4 \pi)^{d\over 2} N \over 2^{2\Delta} \Gamma(\Delta)\Gamma\( 1+\Delta-{d \over 2} \)}
\int_{0}^\infty d\omega ~ {|\lambda(\omega)|^2 \over \omega^{d- 2\Delta-1}}~.
\label{Eden}
\eea
This formula combined with conformal spectral function calculated in Appendix \ref{cftspec} agrees with \reef{density}.

Note that the energy density is manifestly finite and positive. If the unitary bound $d-2\Delta\leq 2$ is satisfied, both the integrand and the numerical pre-factor are positive.\footnote{One can also reverse this argument and argue that positivity of the total energy results in the unitary bound.} Moreover, for smooth $\lambda(t)$ with compact support the integral over $\omega$ converges at both ends. When the unitary  is saturated  (e.g.~deformation by the scalar mass operator in a free theory)  the logarithmic divergence at the lower bound of the integral is compensated by vanishing numerical pre-factor, and we end up  with finite and positive answer. Of course, \reef{Eden} is only applicable provided that the theory is conformal or $\Delta>d/2$ and $\lambda(\omega)$ is sufficiently broad to ensure transitions into the high energy states where the theory is described by a UV CFT. 

To summarize, the results  presented here  recover the scaling behavior found in  \cite{Buchel:2012gw,Buchel:2013lla,Buchel:2013gba,Das:2014jna,Das:2014hqa,Das:2015jka}, see also \cite{Berenstein:2014cia}. We also calculated the numerical coefficient in front of the scaling factor for a generic CFT. Of course, this coefficient depends on the shape of the pulse. Provided $\delta \lambda$ is small enough the higher order corrections in $\delta\lambda$  are suppressed by at least one power of  $\delta\lambda\delta t^{d-\Delta}\ll 1$ relative to the leading order result \reef{Eden}.

\section{Commutator}
\label{intcom}

In this Appendix we calculate the vacuum expectation value of the commutator of two primaries $\mathcal{O}(x)$ having conformal weight $\Delta$. Based on the Euclidean correlator
\be
\langle0| \mathcal{O}(x), \mathcal{O}(y)|0\rangle_\mt{E}={N\over |x-y|^{2\Delta}}~,
\ee
the following relation holds in Lorentzian time\footnote{We use translational invariance to set one of the insertion point at the origin.}
\beq
 \langle 0| [\mathcal{O}(t,\vec x), \mathcal{O}(0) ] |0\rangle = {N\over \big( -(t-i\epsilon)^2 + \vec x^2 \big)^\Delta} -{N\over \big( -(t+i\epsilon)^2 + \vec x^2 \big)^\Delta} ~.
\eeq
Defining now the interval between the insertion points $s^2\equiv-t^2+\vec x^{\,2}$, we rewrite it as follows
\beq
 \langle 0| [\mathcal{O}(t,\vec x), \mathcal{O}(0) ] |0\rangle = \Bigg[{N\over \big( s^2+i\epsilon \big)^\Delta} -{N\over \big( s^2 - i \epsilon \big)^\Delta}\Bigg] \Big[\theta(t)-\theta(-t)\Big]
 \label{comm}
\eeq
Obviously, this expression vanishes in the limit $\epsilon\to 0$ if $s^2>0$. For $s^2<0$, the commutator is readily evaluated if one substitutes 
\beq
 \lim_{\epsilon\to 0}\big(-|s^2| \pm i \epsilon\big) = |s^2| \exp(\pm i \pi)
\eeq
into (\ref{comm}). Hence,
\bea
 \langle 0| [\mathcal{O}(t,\vec x), \mathcal{O}(0) ] |0\rangle &=&  - 2Ni\, {\sin(\pi\Delta)\over (-s^2)^\Delta}\,\Theta(-s^2) ~ \text{sign}(t)~, 
  \label{comm1}
\eea
where $\Theta(x)$ is the standard step function that equals 1 for $x>0$ and vanishes for negative $x$. 

Note that the point $s^2=0$ should be treated in the distributional sense. If, for instance, $\Delta=n$ is an integer, then \reef{comm1} vanishes identically unless $s^2= 0$. In particular, starting from (\ref{comm}) one can use the equality between the distributions
\beq
 \lim_{\epsilon\to 0}{1\over z\pm i\epsilon} = \text{P}{1\over z} \mp i\pi\delta(z)~.
\eeq
to show that for integer $\Delta=n$
\beq
  \langle 0| [\mathcal{O}(t, \vec x), \mathcal{O}(0) ] |0\rangle = 2 \pi N i {(-1)^{n} \over \Gamma(n)} ~ \delta^{(n-1)}(s^2)~ \text{sign}(t) \quad ,
  \label{comm2}
\eeq
where $n-1$ derivatives of the delta function are taken with respect to its argument. It is instructive to show that generic expression (\ref{comm1}) agrees with (\ref{comm2}) in the limit $\Delta\to n$, and we illustrate it now.

First, we integrate (\ref{comm1}) over spatial directions
\beq
 \int d^{d-1}\vec x \, \langle 0| [\mathcal{O}(t, \vec x), \mathcal{O}(0) ] |0\rangle = -  2N i\, \sin(\pi\Delta)~ \text{sign}(t)
 \int d^{d-1}\vec x ~\Theta(-s^2) {1\over (-s^2)^\Delta } ~.
\eeq
In spherical coordinates, we have 
\beq
\int d^{d-1}\vec x ~ \Theta(-s^2) {1\over (-s^2)^\Delta } = 
{2 \pi^{d-1\over 2} \over \Gamma\big( {d-1\over 2} \big)} \int_0^t dr \, r^{d-2} ~ {1\over (t^2-r^2)^\Delta} = { \pi^{d-1\over 2} \Gamma(1-\Delta)\over \Gamma\big( {d-2\Delta+1\over 2} \big)} |t|^{d-2\Delta-1}~.
\eeq
Hence, for general $\Delta$
\beq
 \int d^{d-1}\vec x \, \langle 0| [\mathcal{O}(t, \vec x), \mathcal{O}(0) ] |0\rangle = -  iN
 {2 \pi^{d-1\over 2} \Gamma(1-\Delta)\sin(\pi\Delta)\over \Gamma\big( {d-2\Delta+1\over 2} \big)} ~ \text{sign}(t) \,|t|^{d-2\Delta-1}~.
\eeq
or equivalently, using the identity $\Gamma(1-z)\Gamma(z)=\pi/\sin(\pi\,z)$, we finally obtain
\beq
 \int d^{d-1}\vec x \, \langle 0| [\mathcal{O}(t, \vec x), \mathcal{O}(0) ] |0\rangle = -  iN
 {2 \pi^{d+1\over 2} \over \Gamma\big( {d-2\Delta+1\over 2} \big)\Gamma(\Delta)} ~ \text{sign}(t) \,|t|^{d-2\Delta-1}~.
 \label{icomm}
\eeq

Next, we use (\ref{comm2}) to repeat the same calculation for a particular case of integer $\Delta=n$. This time we have
\beq
 \int d^{d-1}\vec x \, \langle 0| [\mathcal{O}(t, \vec x), \mathcal{O}(0) ] |0\rangle = - {2 \pi i N\over \Gamma(n) }
 ~ \text{sign}(t)\Big({d\over dt^2}\Big)^{n-1}\int d^{d-1}\vec x ~ \delta(\vec x^{\,2} - t^2) ~.
\eeq
In spherical coordinates
\beq
\int d^{d-1}\vec x ~ \delta(\vec x^{\,2} - t^2) = 
{2 \pi^{d-1\over 2} \over \Gamma\big( {d-1\over 2} \big)} \int_0^\infty dr \, r^{d-2} ~ {\delta(r - t) + \delta(r+t)\over 2|t|}= { \pi^{d-1\over 2} \over \Gamma\big( {d-1\over 2} \big)} |t|^{d-3}~.
\eeq
Thus we get for integer $\Delta=n$
\beq
 \int d^{d-1}\vec x \, \langle 0| [\mathcal{O}(t, \vec x), \mathcal{O}(0) ] |0\rangle = -iN{2 \pi^{d+1\over 2} \over   \Gamma\big( {d-2n+1\over 2} \big)\Gamma(n)} ~ \text{sign}(t)
 \,|t|^{d-2n-1} ~ .%~ \text{for} ~ d\geq 3~.
 \label{master}
\eeq
The above expression is in full agreement with the general formula (\ref{icomm}). This completes the proof that the general distribution (\ref{comm1}) converges to (\ref{comm2}) in the limit $\Delta\to n$.

\section{Master integral \reef{I_1} }
\label{masI_1}

In this Appendix we evaluate \reef{I_1}. Our main tool is the Mellin-Barnes (MB) representation\footnote{The real constant $c$ is chosen such that the integration contour separates the left and right series of poles of the gamma functions occuring in the integrand.}
\be
 {1\over \Big(A^{\,2} - M^{\,2}\Big)^\nu}={1\over \Gamma(\nu)}\, {1\over 2\pi i} \int_{c-i\infty}^{c+i\infty} ds ~  {(-M^2)^s\over (A^{\,2})^{\nu+s}} ~ \Gamma(-s) ~ \Gamma(\nu+s)
 ~, \quad\quad -\nu<c<0 ~.
 \label{MB}
\ee

We start from shifting the integration variable $\vec x_1\to \vec x_1+\vec x_2$ and using the MB for the first two terms in the denominator of the integrand in \reef{I_1}
\bea
I_1&=&{-1\over  (2\pi)^2 \Gamma^2\({\Delta_i\over 2}\)} 
\int_{c-i\infty}^{c+i\infty} ds_1 ~ \Gamma(-s_1) ~ \Gamma(\Delta_i/2+s_1) \int_{c-i\infty}^{c+i\infty}  ds_2 ~ \Gamma(-s_2) ~ \Gamma(\Delta_i/2+s_2) 
\non
&&\times \int d^{d-1}\vec x_1 \int d^{d-1}\vec x_2 \, 
{(-T_1^2)^{s_1}(-T_2^2)^{s_2}\over  |\vec x_1+\vec x_2|^{\Delta_i+2s_1} |\vec x_2|^{\Delta_i+2s_2} \big( \vec x_1^{\,2} - T^2\big)^{2\Delta - \Delta_i\over 2}} ~.
\eea
Next we integrate over $\vec x_2$ and $\vec x_1$ with integral over $\vec x_2$ being done first,
\bea
I_1&=&{-\pi^{d-1}\over  (2\pi)^2 \Gamma\({d-1\over 2}\)} \big(-T^2\big)^{d-1-{\Delta_i\over 2}  - \Delta}
\int_{c-i\infty}^{c+i\infty} ds_1    \int_{c-i\infty}^{c+i\infty}  ds_2 \, z_1^{s_1} \, z_2^{s_2}
\non
&&\times ~
 {\Gamma(-s_1)  \Gamma(-s_2)  \Gamma\( \Delta_i + s_1 + s_2 -{d-1\over2}\)\Gamma\( {d-1-\Delta_i\over2}-s_1\)\Gamma\( {d-1-\Delta_i\over2}-s_2\)
 \over \Gamma^2\({\Delta_i\over 2}\)  \Gamma\({2\Delta - \Delta_i\over 2}\)}
\non
&&\times ~ \Gamma\( {\Delta_i\over 2}+\Delta +s_1+s_2-d+1 \) \, 
~,
\eea
where we introduced two dimensionless parameters $z_i=T_i^2/T^2$, and the following master integrals have been used\footnote{Two integrals \reef{2int} can be evaluated using the common technique of Feynman parametrization.}
\bea
 \int d^{d-1}\vec x \, {1\over |\vec x+\vec x_1|^{2\alpha} |\vec x|^{2\beta}}&=&
 \pi^{d-1\over 2} \, { \Gamma\( \alpha+\beta -{d-1\over2}\)\Gamma\( {d-1\over2}-\alpha\)\Gamma\( {d-1\over2}-\beta\)\over \Gamma(\alpha)\Gamma(\beta)\Gamma(d-1-\al-\bt)} 
\( \vec x^{\,2}_1 \)^{{d-1\over 2}-\alpha-\beta} ~,
 \non
  \label{2int}
 \\
 \int  { d^{d-1}\vec x \over \( \vec x^{\,2}-T^2\)^{\alpha} |\vec x|^{2\beta}}&=&
 \pi^{d-1\over 2} { \Gamma\( \alpha+\beta -{d-1\over2}\)\Gamma\( {d-1\over2}-\beta\)\over \Gamma\({d-1\over 2}\) \Gamma(\al)}\, \big(-T^2\big)^{{d-1\over 2} -\al-\bt} ~.
\nonumber
\eea

Taking $t$ sufficiently large results in $|z_1|>1$. In this case we should close the $s_1$-contour to the left 
\bea
I_1\Big|_{|z_1|>1}&=&{\pi^{d-1}\over  2\pi i \, \Gamma\({d-1\over 2}\)} \big(-T^2\big)^{d-1-{\Delta_i\over 2}  - \Delta}
  \sum_{n=0}^{\infty} {(-1)^n \over \Gamma(n+1)} z_1^{{d-1\over2} -\Delta_i-n}\int_{c-i\infty}^{c+i\infty}  ds_2 \,  \({z_2\over z_1}\)^{s_2}
\non
&&\times ~
 {\Gamma\( \Delta_i + n + s_2 -{d-1\over2}\)  \Gamma(-s_2)   \Gamma\( {\Delta_i\over2} + n + s_2 \)
\Gamma\( {d-1-\Delta_i\over2}-s_2\)
 \over \Gamma^2\({\Delta_i\over 2}\)  \Gamma\({2\Delta - \Delta_i\over 2}\)}
\non
&&\times ~ \Gamma\( \Delta -n -{d-1\over 2}-{\Delta_i\over 2} \) \, 
\non
&+&
{\pi^{d-1}\over  2\pi i \, \Gamma\({d-1\over 2}\)} \big(-T^2\big)^{d-1-{\Delta_i\over 2}  - \Delta}
  \sum_{n=0}^{\infty} {(-1)^n \over \Gamma(n+1)} z_1^{d-1 -{\Delta_i\over 2}-\Delta-n}\int_{c-i\infty}^{c+i\infty}  ds_2 \,  \({z_2\over z_1}\)^{s_2}
\non
&&\times ~
 {\Gamma\( {\Delta_i\over 2}+\Delta+n +s_2-d+1\)  \Gamma(-s_2)   \Gamma\(\Delta -{d-1\over2} + n + s_2 \)
\Gamma\( {d-1-\Delta_i\over2}-s_2\)
 \over \Gamma^2\({\Delta_i\over 2}\)  \Gamma\({2\Delta - \Delta_i\over 2}\)}
\non
&&\times ~ \Gamma\( {\Delta_i\over 2} -n -\Delta + {d-1\over 2} \)~.
\eea
It follows from \reef{1ptO2} that we only need to consider the range $t_1>t_2$, \ie when $|z_2|>|z_1|$. In this range the $s_2$-contour should be closed to the left, and we obtain
\bea
&&\scalebox{1.0}{$
I_1\Big|_{t_1>t_2, \,|z_1|>1}={\pi^{d-1}\over   \, \Gamma\({d-1\over 2}\)} \big(-T^2\big)^{d-1-{\Delta_i\over 2}  - \Delta}
  \sum_{m,n=0}^{\infty} {(-1)^{n+m} \over \Gamma(n+1)\Gamma(m+1)} ~ 
  {  \Gamma\( {\Delta_i\over2} + n + m \)  \Gamma\( \Delta -n -{d-1+\Delta_i\over 2} \) \over 
  \Gamma^2\({\Delta_i\over 2}\)  \Gamma\({2\Delta - \Delta_i\over 2}\)}    
  $}
\non
\non
&&\scalebox{1.0}{$\times ~z_1^{{d-1\over2} -\Delta_i-n}   \[
 \Gamma\( \Delta_i + n + m -{d-1\over2}\)  \Gamma\( {d-1-\Delta_i\over2}-m\) \({z_2\over z_1}\)^{{d-1\over 2} - \Delta_i-n-m} \right.
  $}
\non
\non
&&\scalebox{1.0}{$ \left. + ~\Gamma\(   -{d-1-\Delta_i\over2} -m\)  \Gamma\( {d-1\over2}+n+m\) \({z_2\over z_1}\)^{-{\Delta_i\over 2} -n-m} \]
 $}
 \non
 \label{I1}
 \\
 &&\scalebox{1.0}{$
+{\pi^{d-1}\over   \, \Gamma\({d-1\over 2}\)} \big(-T^2\big)^{d-1-{\Delta_i\over 2}  - \Delta}
  \sum_{m,n=0}^{\infty} {(-1)^{n+m} \over \Gamma(n+1)\Gamma(m+1)} ~ 
  {  \Gamma\( \Delta + n + m - {d-1\over2} \)  \Gamma\( {d-1+\Delta_i\over 2} - \Delta -n  \) \over 
  \Gamma^2\({\Delta_i\over 2}\)  \Gamma\({2\Delta - \Delta_i\over 2}\)}    
  $}
  \non
\non
&&\scalebox{1.0}{$\times ~z_1^{d-1 -{\Delta_i\over 2}-\Delta-n}   
\[\Gamma\(   {d-1-\Delta_i\over2} -m\)  \Gamma\( {\Delta_i\over2}+\Delta+n+m-d+1\) \({z_2\over z_1}\)^{d-1-\Delta-{\Delta_i\over 2} -n-m} \right.
  $}
\non
\non
&&\scalebox{1.0}{$ \left. + ~  \Gamma\( \Delta + n + m -{\Delta_i\over2}\)  \Gamma\( {\Delta_i-d+1\over2}-m\) \({z_2\over z_1}\)^{{d-1\over 2} - \Delta-n-m} \]~,
 $}
\eea
The double sum in this expression is known as Appell's hypergeometric function of two variables
\be
 F_4(a,b;\,c,d;\, x,y)\equiv\sum_{m,n=0}^{\infty} {x^m\over m!}\, {y^n\over n!}\, {(a)_{m+n} (b)_{m+n} \over (c)_m (d)_n} \,
 \label{Appell}
\ee
where $(a)_m=\Gamma(a+m)/ \Gamma(a)$. Thus, we can rewrite $I_1$ in terms of linear combination of four Appell's hypergeometric functions\footnote{We used the following identity 
\be \Gamma(x-n)={\Gamma(x)\Gamma(1-x)\over (-1)^n \Gamma(1-x+n)}\, , \quad n\in \mathbb{N} \ee to match \reef{Appell} with various terms in \reef{I1}. }
\bea
%\scalebox{1.0}{$
 I_1&=&\big(-T^2\big)^{d-1-{\Delta_i\over 2}  - \Delta} 
 {\pi^{d-1}  \over \Gamma\({d-1\over 2}\) \Gamma^2\({\Delta_i\over 2}\) \Gamma\( {2\Delta-\Delta_i\over 2}\)} 
 % $}
 \non \non
&&\times\Bigg[  \Gamma\Big({2\Delta-d+1-\Delta_i \over 2} \Big) \Gamma\Big( {d-1-\Delta_i\over 2} \Big) \Gamma\Big({2\Delta_i-d+1 \over 2} \Big)\Gamma\({\Delta_i\over 2}\)
 \non \non
&& %\scalebox{1.0}{$
 \times ~  z_2^{d-1-2\Delta_i\over2}
F_4\({\Delta_i\over 2}, {2\Delta_i - d +1\over 2};\, {3+\Delta_i-d\over 2},\, {d+1+\Delta_i-2\Delta\over 2};\, {z_1\over z_2}, {1\over z_2} \)
 %$}
 \non\non
 &&%\scalebox{1.0}{$
+  \Gamma\Big({2\Delta-d+1-\Delta_i \over 2} \Big) \Gamma\Big( {d-1-\Delta_i\over 2} \Big) \Gamma\Big({d-1 \over 2} \Big) \Gamma\({\Delta_i\over 2}\)
 \non\non
 &&\times~ 
 z_1^{d-1-2\Delta_i\over2} \({z_1\over z_2}\)^{\Delta_i\over 2}
  F_4\({\Delta_i\over 2}, {d-1\over 2};\, {d+1-\Delta_i\over 2},\, {d+1+\Delta_i-2\Delta\over 2};\, {z_1\over z_2}, {1\over z_2} \)
 %$}
\non
% \label{I1fin}
\non
 &&%\scalebox{1.0}{$
+  \Gamma\Big({2\Delta-d+1\over 2} \Big) \Gamma\Big( {d-1-\Delta_i\over 2} \Big) \Gamma\Big({d-1 +\Delta_i\over 2}-\Delta \Big) \Gamma\({\Delta_i\over 2}+\Delta-d+1\)
 \non\non
 &&\times~ 
 z_2^{d-1-\Delta- {\Delta_i\over2} } 
  F_4\({\Delta_i\over 2}+\Delta-d+1, \Delta-{d-1\over 2};\, {3-d+\Delta_i\over 2},\, {3-d-\Delta_i\over 2};\, {z_1\over z_2}, {1\over z_2} \)
 %$}
 \non\non
 && 
+  \Gamma\Big({d-1-2\Delta+\Delta_i \over 2} \Big) \Gamma\Big( {\Delta_i+1-d\over 2} \Big) \Gamma\Big(\Delta-{d-1 \over 2} \Big) \Gamma\(\Delta-{\Delta_i\over 2}\)
 \non\non
 &&\scalebox{1.1}{$\times~ 
 z_1^{d-1-{\Delta_i\over2}-\Delta} \({z_2\over z_1}\)^{{d-1\over 2} - \Delta}
  F_4\(\Delta-{\Delta_i\over 2}, \Delta-{d-1\over 2};\, {d+1-\Delta_i\over 2},\, {3-d-\Delta_i\over 2};\, {z_1\over z_2}, {1\over z_2} \) \Bigg] ~.
 $}
\non\nonumber
 \eea
Note that we suppressed the restriction $t_1>t_2, \,|z_1|>1$ in \reef{I1fin} since other values are treated by analytic continuation.

\section{Spatial correlator without MB}
\label{noMB}

In this Appendix we study quenched spatial correlator without use of the MB representation. The calculations presented here clearly illustrate the advantage of MB method used in the text. First, we introduce $s^2_1\equiv-t^2+(\vec x - \vec y)^{\,2}$ and $s^2_2\equiv-t^2+\vec y^{\,2}$ to rewrite \reef{comm3O} as follows
\bea
 &&\langle 0| [\mathcal{O}_i(t,\vec x)\mathcal{O}_j(t,0), \mathcal{O}_k(0, \vec y) ]  |0\rangle =  { \text{sign}(t) \over |\vec x|^{\Delta_{ijk}}}
\\
&&\times
   ~\Bigg[{C_{ijk}\over \big( s^2_1+i\epsilon \big)^{\Delta_{kij}\over 2}\big( s^2_2+i\epsilon \big)^{\Delta_{kji}\over 2}} 
  -{C_{ijk}\over \big( s^2_1-i\epsilon \big)^{\Delta_{kij}\over 2}\big( s^2_2-i\epsilon \big)^{\Delta_{kji}\over 2}}\Bigg] ~.
    \nonumber
\eea
Taking the limit $\epsilon\to 0$, we arrive at
\bea
 &&\langle 0|  [\mathcal{O}_i(t,\vec x)\mathcal{O}_j(t,0), \mathcal{O}_k(0, \vec y) ] |0\rangle =  
 - {2\,i\, C_{ijk} \sin(\pi\Delta_k)\over \big( - s^2_1 \big)^{\Delta_{kij}\over 2}\big( -s^2_2 \big)^{\Delta_{kji}\over 2}} 
 ~ { \text{sign}(t) \over |\vec x|^{\Delta_{ijk}}} ~\Theta(-s^2_1) \Theta(-s^2_2)
 \nonumber
  \\
 &&  - { 2\,i\, \text{sign}(t) \, C_{ijk}\over |\vec x|^{\Delta_{ijk}}}
 \Bigg({\sin\big({\pi\Delta_{kij}\over 2}\big)\over  \big( - s^2_1 \big)^{\Delta_{kij}\over 2}\big( s^2_2 \big)^{\Delta_{kji}\over 2}}
 \Theta(-s^2_1) \Theta(s^2_2) + (i\leftrightarrow j\,, 1\leftrightarrow2) \Bigg) ~.
\eea

Substituting this expression into (\ref{corr}) and considering first $|\vec x|>2|t-t'|$ for all $t'$ within the interval of order $\delta t$ around $t'=0$, results in the following linear correction to a pure CFT two-point function\footnote{Recall that $\lambda(t')$ vanishes sufficiently fast outside a finite interval of order $\delta t$ around the origin, therefore this region only contributes to the integral over $t'$.}\bea
&& \delta^{(1)}\langle \mathcal{O}_i(t,\vec x)\mathcal{O}_j(t,0)\rangle = 
- 2\, C_{ijk} \, { \sin\big({\pi\Delta_{kji}\over 2}\big) \over |\vec x|^{\Delta_{ijk}}}
\nonumber \\
 &&
 \times \int_{-\infty}^t dt' \lambda(t')  \int_{|\vec y|<|t-t'|} d^{d-1}\vec y 
 \, {1\over  \big(-(t-t')^2+(\vec y-\vec x)^2\big)^{\Delta_{kij}\over 2} \big((t-t')^2-y^2\big)^{\Delta_{kji}\over 2} } 
 \nonumber\\
 &&+(i\leftrightarrow j)
 \nonumber\\
 &&=- 2\, C_{ijk} \, {\sin\big({\pi\Delta_{kji}\over 2}\big) \over |\vec x|^{\Delta_{ijk}}}
   {2\pi^{d-2\over 2}\over \Gamma\big( {d-2\over 2} \big)}\int_{-\infty}^t dt' \lambda(t')  
 \int_0^{t-t'} dy \, y^{d-2} 
 \nonumber \\
 &&\quad\quad\times \int_0^\pi d\theta \sin^{d-3}\theta
 \, {1\over \big(y^2+x^2 - 2yx\cos\theta -(t-t')^2 \big)^{\Delta_{kij}\over 2}  \big((t-t')^2-y^2\big)^{\Delta_{kji}\over 2} } ~,
   \nonumber\\
 &&+(i\leftrightarrow j)
\eea
where in the second equality we introduced the standard spherical coordinates around $\vec x$. 

Now we expand the integrand in the limit $x\gg  t$ and $x\gg \delta t$ to carry out the remaining integrals. The leading and next-to-leading terms are given by
\bea
&& \delta^{(1)}\langle \mathcal{O}_i(t,\vec x)\mathcal{O}_j(t,0)\rangle\Big|_{ x\gg t,\delta t}
= {- 2 \pi^{d+1\over 2} \over \Gamma\big( {\Delta_{kji}\over 2} \big)\Gamma\big({d-\Delta_{kji}+1\over 2}\big)} ~{C_{ijk} \over |\vec x|^{2\Delta_i}}  
\label{2plargeX}
  \\
  &&
  \times\int_{-\infty}^t dt' \lambda(t') (t-t')^{d-\Delta_{kji}-1}
  \Big(1 + {2+\Delta_i-\Delta_j\over d-\Delta_{kji}+1} {(t-t')^2\over |\vec x|^2}+ \ldots\Big)
     +(i\leftrightarrow j)~ . %\quad \text{for} \quad |\vec x|\gg  |t-t'|~.
 \nonumber
\eea
This expression is free of the logarithmic divergences unless $t$ is within the support of $\lambda(t')$ and $\Delta_{kji}\geq d$ and/or $\Delta_{kij}\geq d$ are integers. At late times, $t\gg\delta t$, or equivalently for sufficiently fast but smooth quenches, \reef{2plargeX} is finite and approaches zero. However, at early times this is no longer true. To illustrate this point explicitly let us choose $t=0$, then \reef{2plargeX} takes the same form as \reef{2pscaling}
\bea
&& \delta^{(1)}\langle \mathcal{O}_i(0,\vec x)\mathcal{O}_j(0,0)\rangle\Big|_{ x\gg \delta t}
= {- 2 \pi^{d+1\over 2} \over \Gamma\big( {\Delta_{kji}\over 2} \big)\Gamma\big({d-\Delta_{kji}+1\over 2}\big)} ~{C_{ijk} \over |\vec x|^{2\Delta_i}}   
{\delta\lambda \over (\delta t)^{\Delta_{kji}-d} }
%\label{2pscaling}
  \\
  &&
  \times \int_{-\infty}^0 d\xi f(\xi) (-\xi )^{d-\Delta_{kji}-1} \Big(1+\mathcal{O}\big(\delta t/ x\big)^2 \Big)
     +(i\leftrightarrow j)~ , 
      \nonumber
\eea
where $\xi=t'/\delta t$ is a dimensionless parameter.

\bibliographystyle{utcaps}

\bibliography{quen_lib}

\end{document}